\documentclass[preprint,12pt]{elsarticle}

\usepackage{amssymb}
\usepackage{amsmath}
\usepackage{caption}
\usepackage{subcaption}
\usepackage{xcolor}
\usepackage{comment}

\usepackage{lineno}
\usepackage{multirow}
\usepackage{tabularx,booktabs}
\usepackage{enumitem}
\usepackage{url}
\newcolumntype{Y}{>{\centering\arraybackslash}X}

\begin{document}

\begin{frontmatter}



\title{What Can We Learn from the Travelers Data in Detecting Disease Outbreaks-- A Case Study of the COVID-19 Epidemic}


\author{Le Bao, Ying Zhang, Xiaoyue Niu}

\address{Department of Statistics, Penn State University\\
University Park, PA 16802}

\begin{abstract}

\noindent \textbf{Background:} Travel is a potent force in the emergence of disease. We discussed how the traveler case reports could aid in a timely detection of a disease outbreak. 

\noindent \textbf{Methods:} Using the traveler data, we estimated a few indicators of the epidemic that affected decision making and policy, including the exponential growth rate, the doubling time, and the probability of severe cases exceeding the hospital capacity, in the initial phase of the COVID-19 epidemic in multiple countries. We imputed the arrival dates when they were missing. We compared the estimates from the traveler data to the ones from domestic data. We quantitatively evaluated the influence of each case report and knowing the arrival date on the estimation. We developed a simple disease detection criterion that could help make future decisions.

\noindent \textbf{Findings:} Using only the travel report data, we estimated the travel origin's daily exponential growth rate in a moving window fashion that mimic the reality, and examined the date from which the growth rate was consistently above 0.1 (equivalent to doubling time $<$ 7 days). We found those dates were very close to the dates that critical decisions were made such as city lock-downs and national emergency announcement. In addition, our estimated probability of severe cases exceeding the hospital capacity hit above 0.9 on the day Wuhan announced lock-down. Using only the traveler data, if the assumed epidemic start date was relatively accurate and the traveler sample was representative of the general population, the growth rate estimated from the traveler data was consistent with the domestic data. We also discussed situations that the traveler data could lead to biased estimates. From the data influence study, we found more recent travel cases had a larger influence on each day's estimate, and the influence of each case report got smaller as more cases became available. We provided the minimum number of exported cases needed to determine whether the local epidemic growth rate was above a certain level, and developed a user-friendly Shiny App to accommodate various scenarios.  

\noindent \textbf{Interpretation:} The traveler data are useful information that help the early detection of a disease outbreak in the travel origin. The traveler data also have limitations that would need further information to refine. We advocate that countries should work in a collaborative way to share the traveler information about the travel dates and more detailed travel history at sub-national level, in a timely manner.

 
\noindent \textbf{Funding:} NIH/NIAID 5R01AI136664
\end{abstract}


\end{frontmatter}


\section{Introduction}
Since 2014, there have been five ``public health emergency of international concern" (PHEIC) declarations including the most recent outbreak of COVID-19 \cite{WHO2020Jan31}. Major challenges in early warning and rapid response to an emerging disease outbreak include the lack of epidemiological and laboratory techniques where disease started, and the fear of the negative impact that reporting the outbreak would have on trade and tourism \citep{Heymann2001hot,Morse2007global}. At the same time, travel is a potent force in the emergence of disease \citep{Wilson1995travel}. Travelers play a key role in a disease spreading out from local to global. 


We believe that the case reports among travelers are informative data source for detecting the disease outbreaks. The WHO International Health Regulations (IHR) provided the instructions for nations to report diseases found in incoming travelers, and the regulations were revised to accommodate warnings of unknown infectious diseases \citep{Morse2007global}. \citep{Wu2020nowcasting} estimated the outbreak size of 2019-nCoV in Wuhan from the number of confirmed cases that have been exported to cities outside mainland China shortly after the lockdown in Wuhan. Other early studies that estimated the outbreak size in Wuhan using the cases exported include \cite{Imai2020report2, Chinazzi2020preliminary}. 

In this article, we used the travelers data to estimate two important quantitative measures, the exponential growth rate (equivalent to doubling time) and the probability of severe cases exceeding hospital capacity, to detect the disease outbreak in five travel origins (origins), Wuhan (China), Egypt, Iran, Italy, and United States. We investigated the contributions and limitations of traveler data on understanding the COVID-19 epidemic at its beginning stage. Our main contributions to the literature include: 1. Instead of using accumulated traveler cases up to a certain time point to get one estimate, we provided daily estimates so that we could see how the estimates of the key measures were evolving in the early stage. 2. A lot of the traveler case reports only provided the dates when travelers tested positive, without their actual arrival dates. We demonstrated that the arrival date was important in estimating those key measures and provided a probability model to impute the arrival date. 3. We used a statistically rigorous way to evaluate the influence of each case report and knowing the arrival dates on estimating the exponential growth rate. 4. We used the travel origin's domestic data to benchmark the estimates from the travel case report and pointed out that there was a potential sampling bias in the travelers population so that the prevalence estimated from the traveler data might be different from the domestic estimate. We made the first attempt to adjust the bias but also acknowledge that we would need further information about the difference between travelers and the general population to fully address this issue. 5. We provided a simple criterion to determine the severity of the local epidemic based on the cumulative number of exported cases, and developed an easy-to-use app for practical usage. Finally, we discussed how to enhance global infectious disease surveillance by utilizing traveler case reports in a collaborative way.



\section{Methods}


\subsection{Data Sources}
We chose five places (four countries and one city) whose COVID-19 outbreaks were detected relatively early in their region as illustrative examples of travel origins. They were Wuhan (China), Italy, Iran, Egypt and United States. For each origin, we investigated the exported cases in the early stage of the country's epidemic. We picked the start date as the date when the country announced its first confirmed case. We picked the end date as the date the country announced national emergency or implemented lock-down order. Table \ref{tab:data} summarizes the start date, the end date, the total number of exported traveler case reports and the total number of exported destinations within the study period. Without the daily travel volume, we assumed that the travel volume to each destination was constant over this short period of time. The reasons for using those dates and the associated travel volume data are provided in Appendix A.1.

\begin{table}[!ht]
    \caption{The start date of the local epidemic, the end date of study period, the total number of exported traveler case reports within the study period, and the total number of exported destinations within the study period.}
    \label{tab:data}
    \centering
\resizebox{1\columnwidth}{!}{%
\begin{tabularx}{1.2\textwidth}{c *{4}{Y}}
    \toprule
        & start date of the local epidemic & end date of the study period & number of exported cases & number of destinations\\
        \midrule
      Wuhan & Dec 1, 2019 & Jan 23, 2020 & 11 cases & 7 destinations\\
      U.S. & Jan 20, 2020 & Mar 13, 2020 & 25 cases & 5 destinations\\
      Italy & Jan 31, 2020 & Feb 28, 2020 & 57 cases & 19 destinations\\
      Iran & Feb 19, 2020 & Feb 27, 2020 & 90 cases & 12 destinations\\
      Egypt & Feb 14, 2020 & Mar 6, 2020 & 9 cases  & 4 destinations\\
\bottomrule
\end{tabularx}
}
\end{table}



We obtained the traveler case reports within the period specified above from \cite{xu2020epidemiological} with re-verification of their travel histories using government and media reports. Our analysis focused on the infections that likely occurred in the origin rather than transmitted during the trip. Therefore, we counted multiple confirmed cases from the same tour group or the same family as one, and excluded the cases that were related to the cruise ships because a large number of travelers being infected upon arrival made it difficult to trace the travel history of the first infected case. The travel case report data used in our analysis is provided in Appendix B.




For each travel origin, we also obtained the domestic case reports to model its growth curve as a comparison with the one estimated from the travel report. The city level data for Wuhan were obtained from \citep{Pan2020association}; and the country level data for United States, Italy, Iran and Egypt were obtained from COVID-19 Data Repository by the Center for Systems Science and Engineering (CSSE) at Johns Hopkins University \citep{JHUdata}.


\subsection{Assumptions and key indicators of interest}


In the initial phase of an emerging infectious disease outbreak, the infected cases and the recovered cases are both rare. So, the size of the susceptible population approximately equals to the whole population size, $N$.  Let $\rho_t$ be the prevalence rate (number of infected over total population size) $t$ days after the epidemic first started. We assume that the prevalence rate increases exponentially in this period,
\begin{equation}
\rho_t = \exp{(\beta_0 + \beta_1 t)}, 
\label{eq:rho_t}
\end{equation}
where $\beta_1$ is the exponential growth rate. $N \cdot \exp{(\beta_0)}$ is the number of infected people when the epidemic first started. It came from each country's official announcement of their first confirmed case(s). In our study, this number is either 1 or 2. Note that the exact date of the first infection was hard to know, especially in the beginning of the outbreak, but some preliminary evidence might offer possible dates. We explored different possible start dates in the sensitivity analysis. 

Another commonly used indicator, the doubling time, $T_d$, is given by $T_d = \log2 / \beta_1$. The doubling time is the period that takes the number of infected individuals to double. Both the exponential growth rate ($\beta_1$) and the doubling time ($T_d$) measure the speed of the transmission which determines the scale of the epidemic.

Infection among medical professionals and death rate increase significantly when the health care system is overwhelmed. So, another important indicator we care about is the probability of the number of severe case patients exceeding the hospital capacity. Both the severe case patients and the hospital capacity are not evenly distributed within a country. Therefore, it would not make sense to compare the total number of severe case patients in a country to the national hospital capacity. Instead, we focused on one city level analysis (Wuhan) as an example. 

\subsection{Statistical modeling of the prevalence in the origin country}


Our goal was to use the traveler case reports to infer the prevalence rate, or the exponential growth rate, in the origin country. For each origin, let $n_{it}$ be the number of infected travelers who arrived at destination $i$ at time $t$, and $N_{i}$ be the size of average daily travelers of destination $i$ in the study period. We assumed that the number of infected travelers followed a binomial distribution as below: 

\begin{equation}
n_{it}| N_{i}, \rho_t , \alpha \sim \text{Binomial}(N_{i}, \rho_t \cdot \exp{(\alpha)}),
\label{eq:basic-model}
\end{equation}
where $\rho_t$ was the prevalence rate in the general population, and $\alpha$ captured the potential sampling bias in the travelers. If the group of travelers was a representative sample of the general population, the travelers would have the same rate of infection as the general population, so $\alpha=0$. If for some reason we believed that the travelers were more or less likely to get infected, we would expect a non-zero $\alpha$. 

Instead of using all of the reports up to the end date and providing one estimate of $\beta_1$ for each origin country,  we provided daily update of $\beta_1$ to mimic the real-world decision making process. Specifically, for each origin, starting from the date with the first exported case report, $t_1$, for all $t>t_1$, we ran the above model with all the data up to day $t$ and provide a $\beta_1$ estimate. By doing so, we were able to see how the origin country's $\beta_1$ estimate changed as the exported case reports accumulated, and when we could detect the outbreak with some consistency. 

The destination countries with advanced public health infrastructures were more likely to detect the imported cases, and citizens from those countries might prefer returning to their countries for treatment when feeling sick. Therefore, it was possible that the groups of people who chose to travel to different destinations had different infection rates. We could address this heterogeneity by adding a destination specific random effect parameter $b_i$:
\begin{equation}
n_{it}| N_i, \rho_t , \alpha, b_i \sim \text{Binomial}(N_i,\rho_t \cdot \exp{(\alpha + b_i)}).
\label{eq:heterogeneous-model}
\end{equation}
When fitting the real data, we tested the significance of this heterogeneity parameter $b_i$ and found it not significant because of small sample sizes at the destination level. Therefore, we only reported results from the basic model as in (\ref{eq:basic-model}).

We used Bayesian inference to estimate the parameters. Details about the choice of priors and hyper parameters are provided in Appendix A.2. 

\subsection{Imputing the missing arrival dates}
\label{sec:impute}

In the initial period of the epidemic, there was not an airport-based traveler screening system. Therefore, a non-negligible proportion of arrival dates of the confirmed cases were missing in the early government and media reports. The accumulated missing proportions were 9.1\% for Wuhan (China), 60.0\% for United States, 89.5\% for Italy, 88.9\% for Iran and 66.7\% for Egypt. 
To accurately estimate $\beta_1$, it was necessary to impute those missing arrival dates by estimating the time interval between the arrival date and case confirmation date for each case. We modeled this time interval $T_{ij}$ for destination $i$ and case $j$ by using a negative binomial distribution with a destination-specific mean $\mu_i$ and a dispersion parameter $\phi$. In addition, some destination countries did not report any arrival dates of infected travelers. To estimate the time intervals for those countries, we pooled $\mu_i$ towards a common mean. Specifically, the model was as follows: 
\begin{eqnarray*}
T_{ij}|\mu_i, \phi &\sim& NegBinomial(\mu_i,\phi)\\
\mu_i &\sim& Gamma(\lambda, 1)
\end{eqnarray*}
Bayesian estimation was implemented and the choice of priors is provided in Appendix A.2. We used the observed pairs of arrival dates and confirmation dates to estimate the negative binomial parameters and generated posterior predictive distribution of the missing $T_{ij}$. 

Finally, due to the delay of the diagnostic confirmation, we expected a proportion of travelers would not have been tested by the end of the study, and the undiagnostic rate would be higher in the later dates. For travelers arriving at destination $i$ on day $t$, we down-scaled the average daily traveler size $N_i$ by a factor of $q_{it}=P(T_{ij} \le t_{\mbox{end}} - t)$, where $t_{\mbox{end}}$ is the study end date and $T_{ij}$ follows the negative binomial distribution. The final model we fitted was the following:

\begin{equation}
n_{it}| N_i, \rho_t , \alpha \sim \text{Binomial}(N_i\cdot q_{it} , \rho_t \cdot \exp{(\alpha)}),
\label{eq:final-model}
\end{equation}

Rather than imputing the missing arrival date with the posterior mean or median of $T_{ij}$, we fitted model (\ref{eq:final-model}) with each posterior sample of $T_{ij}$ and the corresponding $q_{it}$. Therefore, the final uncertainty estimate of $\beta_1$ included the uncertainty of the missing arrival dates.




\subsection{Comparison with the domestic data estimation}

Our goal was to estimate the epidemic in the general population of those five origins by using their travelers data. As pointed out before, using only the travelers data might generate potential biases. We formally investigated the biases by comparing the estimates from the travelers data with the ones from the domestic case reports.

\cite{Hao2020reconstruction} carefully studied and reconstructed the full transmission dynamics in the early period of the COVID-19 epidemic in Wuhan by fitting a 7-compartment model. From their model output, we could derive the exponential growth rate before Jan 23, 2020 (assumed to be a constant) being 0.187 with 95\% credible interval (0.178, 0.196). Unfortunately, similar results were not readily available for the other four travel origins due to limitations of the data and knowledge about their early stages. 
Instead, using the same data as in \cite{Pan2020association}, we fitted an exponential growth model to the domestic daily records of new infections. We found that for Wuhan (China), the directly fitted exponential growth rate was 0.175 with 95\% confidence interval (0.160, 0.190). The large overlap between the two confidence intervals suggested that the exponential growth curve was a good approximation to transmission dynamic in the beginning of the disease outbreak. Therefore, for the other origins, we used the exponential growth rate fitted by this simple model using domestic records in \citep{JHUdata} to serve as a comparison with the one estimated from the travelers data.

\subsection{Estimating the probability of severe case exceeding hospital capacity}

\cite{liang2020clinical} estimated that 19.2\% of infected individuals needed to be hospitalized. The number of hospital beds available to accept severe COVID-19 patients in Wuhan on January 21 was 800, according to \citep{xinhuanet}. We denote the number of severe case patients on day $t$ by $S_t$, and $S_t=0.192 \cdot N \cdot \rho_t$. For Wuhan, the probability that the number of severe cases exceed hospital capacity is then defined as 
\[
    p_t=P(S_t>800)
\]
Using data up to day $t$, the posterior distribution of $\beta_1$ estimated from model (\ref{eq:final-model}) gave us the posterior distribution of $S_t$. From there, we updated $p_t$, the daily probability of hospital being overwhelmed.

\subsection{Evaluating the traveler data impacts}


In decision theory, we can evaluate the impact of a piece of information by comparing the risks with and without that information. The reduction in the risk is referred to as the value of information (VOI). In this study, we evaluated two types of impacts on estimating the exponential growth rate $\beta_1$. They were 1. the traveler cases confirmed on day $t$, from the first date a travel case was confirmed to the end date of the study; and 2. knowing the actual arrival date of each case.

In Bayesian inference, when the goal is to provide a probabilistic estimate of a certain indicator, one proper risk is the integrated quadratic distance (IQD) \citep{Thorarinsdottir2013using}, which measures the difference between two distributions. The IQD can be calculated as follows:
\begin{equation}
r(F,G) = E_{F,G}|\theta_F-\theta_G| - \frac{1}{2}\Big[ E_F|\theta_F-\theta_F'| + E_G|\theta_G-\theta_G'| \Big],
\label{eqn:iqd}
\end{equation}
where, in Bayesian decision theory, $G$ is the estimated distribution and $F$ is the true distribution of $\theta$, $\theta_F$ is a random variable following the distribution $F$, $\theta_G$ is a random variable following the distribution $G$, and $\theta_F$ and $\theta_F'$ are two independent random variables following the distribution $F$. The true distribution is often unknown and thus replaced by the posterior distribution of $\theta$ given the full data. In that case, the VOI of any piece of information reduces to $r(F, G)$, where $F$ is $\theta$'s posterior distribution given the full data, and $G$ is $\theta$'s posterior distribution given the partial data by removing the piece of information we are interested in evaluating, such as missing arrival dates or one day's report \citep{Parsons2018value}. 

For each day after the first case report, we would like to evaluate the influence of all available information on making the current day's decision. Specifically, to evaluate the impact of daily travel case reports on day $t$, we performed the following analysis.
\begin{enumerate}
    \item Estimated the posterior distribution of $\beta_1$ given all the data up to day $t$, denoted as $F_t$.
    \item For $i \leq t$, removed day $i$'s case report from the data, and estimated the posterior distribution of $\beta_1$ given the partial data, denoted by $G^c_{t(i)}$, where $c$ stands for case reports. Calculated $r^c_{t(i)}$ using Equation (\ref{eqn:iqd}), by setting $F=F_t$, and $G=G^c_{t(i)}$.
\end{enumerate}

To evaluate the impact of knowing the arrival date, we performed the similar analysis. The only difference was setting $G=G^a_{t(i)}$ the posterior distribution of $\beta_1$ given the partial data that removed the arrival date of day $i$'s report, and imputed the missing arrival date following the procedure described in Section \ref{sec:impute}. The superscript $a$ stands for arrival dates.

The procedure above allowed us to evaluate the information in three different aspects: 1. For each day's decision, rank the influence of all the previous days' case reports and the arrival dates; 2. For each day's decision, compare the influence of the number of cases v.s. knowing the arrival date for the same time point ($r^c_{t(i)}$ v.s. $r^a_{t(i)}$); and 3. As time moves forward, compare the influence of the same day's report ($r^c_{t(i)}$ v.s. $r^c_{t'(i)}$) and arrival date ($r^a_{t(i)}$ v.s. $r^a_{t'(i)}$). 

Finally, when we remove a day's case report, we would like the effects to be purely due to the number of cases not observed, rather than a mix of not knowing the arrival date and not observing the cases. Therefore, we only performed this analysis on the Wuhan data since Wuhan only had one missing arrival date in our study period while the rest of the origins all had over 50\% of missing arrival dates. 

\subsection{Detecting future disease outbreaks}
Finally, we would like to develop a useful tool for policy makers to determine the severity of a country's epidemic based on the number of exported cases. Statistically, we could form the question as a hypothesis testing problem. On each day, based on the cumulative number of exported cases up to that day, we could test whether the exponential growth rate is significantly above a certain threshold, for example 0.1.
\[
\begin{array}{ccc}
  H_0: \beta_1 = 0.1   & v.s. & H_1: \beta_1 > 0.1
\end{array}
\]
We used a simulation based hypothesis testing procedure as follows:
\begin{enumerate}
    \item Define the initial prevalence at a very low level such as one over the population size of the travel origin. 
    \item Calculate the time series of true prevalence rates following Equation (\ref{eq:rho_t}) given the exponential growth rate at the null ($H_0$) value.
    \item Let $N$ be the daily average of outbound traveler sizes. 
    \item Generate a time series of daily number of infected travelers from the binomial distributions provided in Equation (\ref{eq:basic-model}). Record the cumulative number of traveler cases up to each day since the first exported case. 
    \item Repeat Step 4 100,000 times. Summarize the empirical distributions of the cumulative number of traveler cases over the repetitions in two ways depending on whether the initial date of the local infection is known:
    \begin{enumerate}
        \item If the initial date is known, we summarize the empirical distributions of the cumulative number of traveler cases stratified by the number of days since the initial local infection. 
        \item If the initial date is unknown, we summarize the empirical distributions of the cumulative number of traveler cases stratified by the number of days since the first exported case. In this case, day 1 represents the date when the first exported case happens. 
    \end{enumerate}
    An extremely large number of traveler cases indicates a higher than expected growth rate. 
    \item Define a significance level and visualize the cumulative number of traveler cases that would be needed to reject $H_0$ for each day since the first exported case.
\end{enumerate}

\section{Results}

The results section is organized in the following way. In Section \ref{sec:time-interval} we present the results for the estimated time intervals between arrival date and case confirmation date by origin and destination. In Section \ref{sec:beta1}, we illustrate how the estimated origin exponential growth rate and the doubling time changed as more case reports became available over time, and investigate the probability of severe COVID-19 cases exceeding the number of available beds in Wuhan, China. In Section \ref{sec:sensitivity-initial}, we conduct the sensitivity analysis to the initial number of cases. In Section \ref{sec:VOI}, we discuss the impact of knowing the traveler cases and the travel dates on shaping the epidemic trend. 


\subsection{Time intervals between arrival and case confirmation}
\label{sec:time-interval}

\begin{figure}[!ht]
    \begin{minipage}{0.5\textwidth}
        \caption*{\tiny{(a) Departure from Wuhan, China}}
        \includegraphics[width=\linewidth]{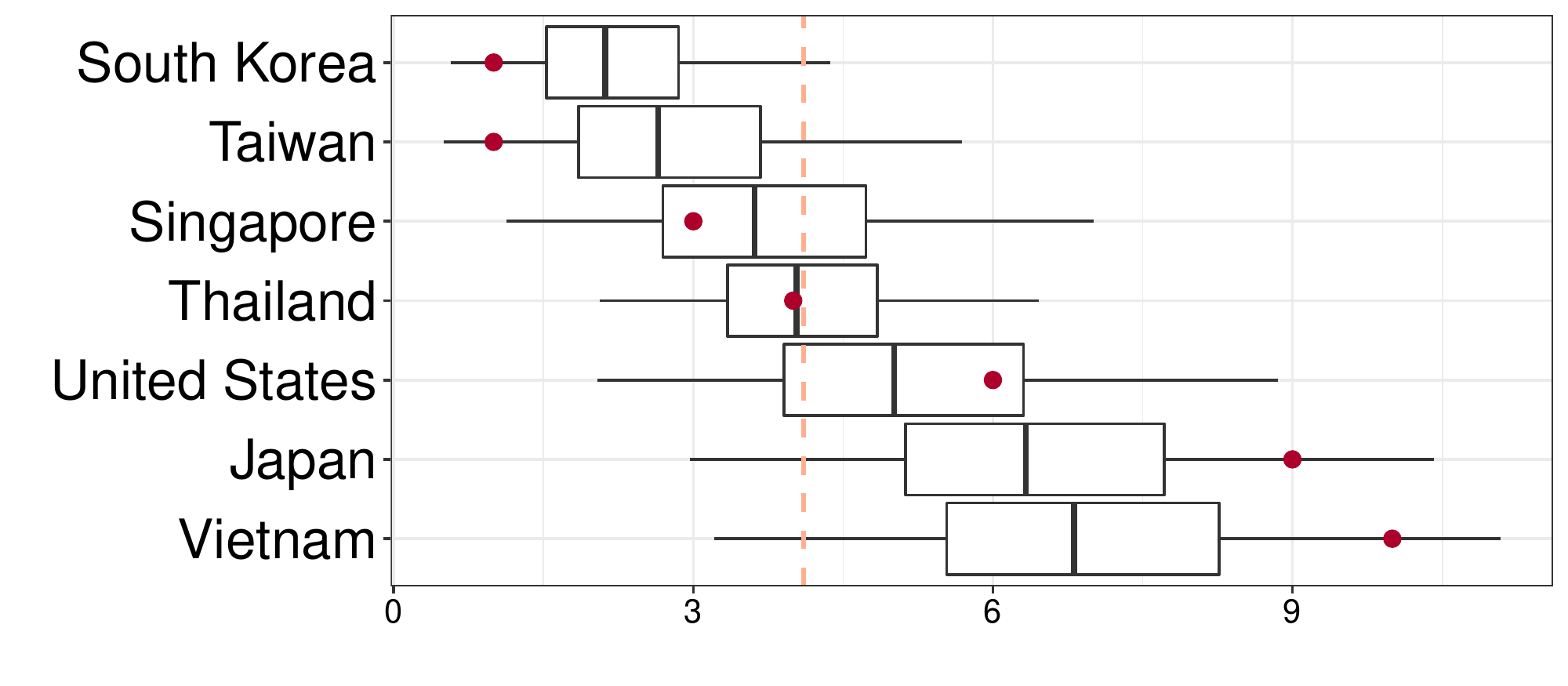}
        \caption*{\tiny{(d) Departure from Iran}}
        \includegraphics[width=\linewidth]{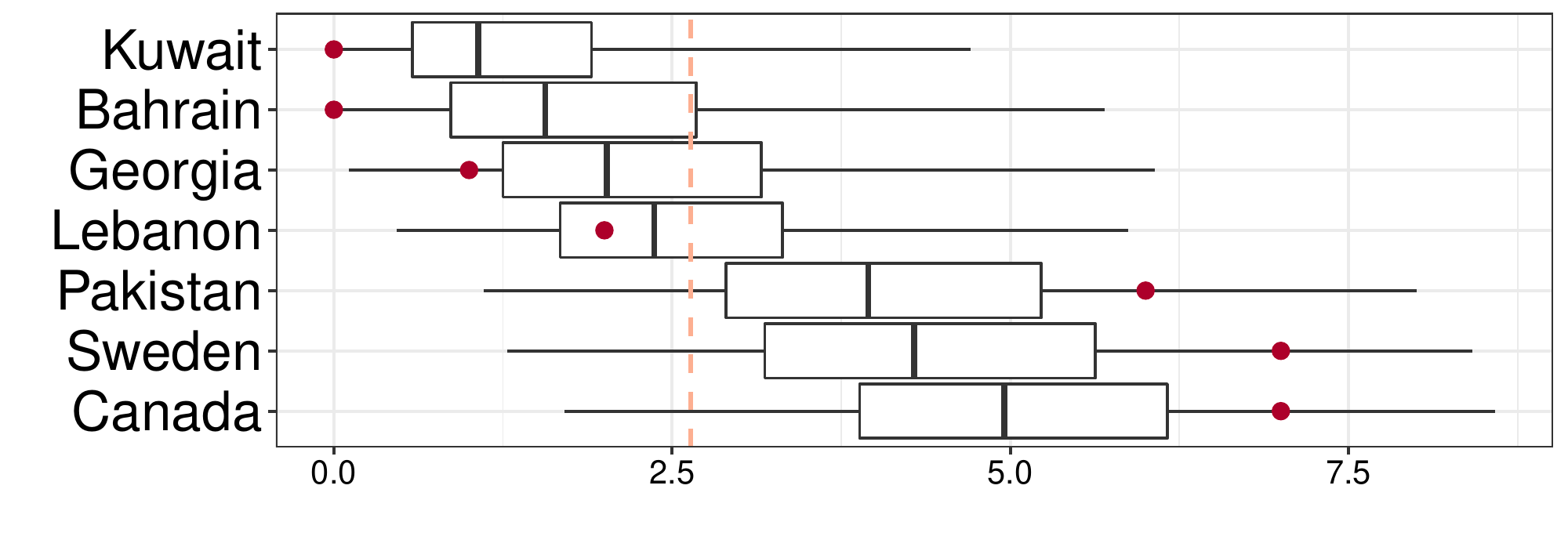}

    \end{minipage}
    \begin{minipage}{0.5\textwidth}
        \caption*{\tiny{(b) Departure from United States}}
        \includegraphics[width=\linewidth]{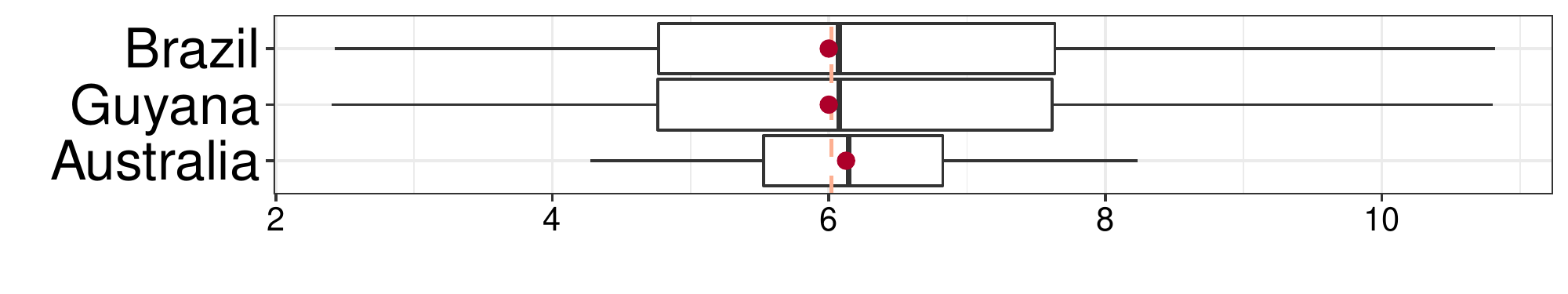}
        \caption*{\tiny{(c) Departure from Egypt}}
        \includegraphics[width=\linewidth]{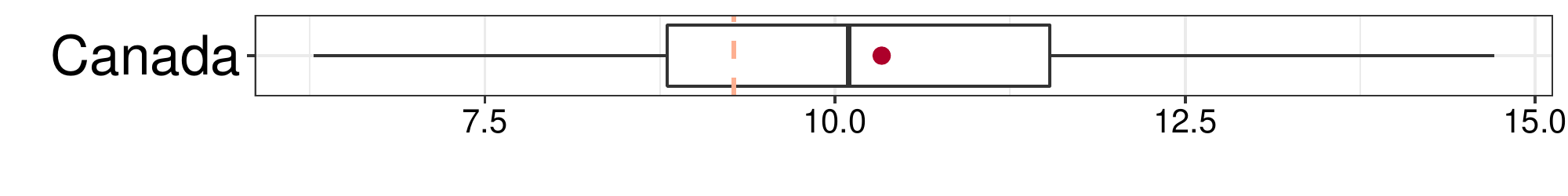}
        \caption*{\tiny{(e) Departure from Italy}}
        \includegraphics[width=\linewidth]{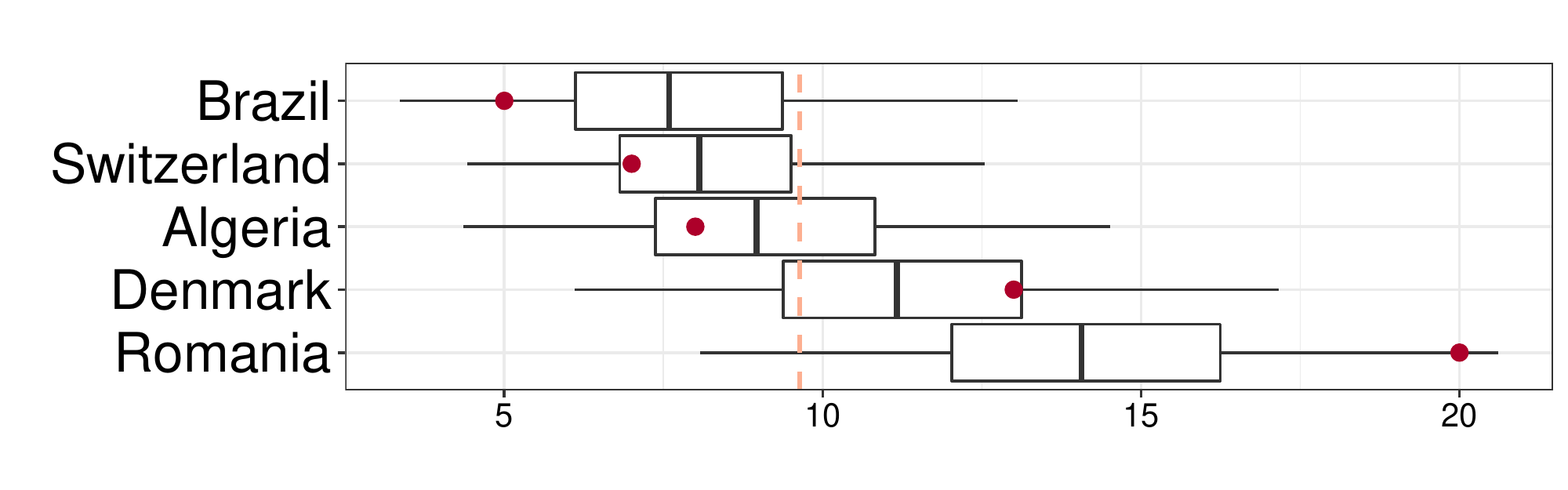}
    \end{minipage}
\newline
    \caption{Posterior median, interquartile, and 95\% credible interval of time to diagnosis by departure city/country and destinations. Each box represents the interquartile of the posterior of time to diagnosis by departure city/country, with the middle bar indicating the posterior median and the whisker showing the 95\% credible interval. The red dots are observed time to diagnosis for the pairs of arrival dates and confirmed dates. The pink dashed line represents the overall median time to diagnosis across all destinations. 
}
\label{post}
\end{figure}

Figure 1 presents the distribution of time interval from arrival to diagnosis by origin: (a) Wuhan, China, (b) United States, (c) Egypt, (d) Iran and (e) Italy. Iran travelers had the shortest time interval with a median of 2.64 days shown by the pink dashed line and a  95\% credible interval of (1.51,  4.20) days. The first exported case from Iran was confirmed on Feb 20, 2020. Iran announced the first two COVID-19 infection related deaths on Feb 19 \cite{world2020coronavirus}, and reported 29 new cases by Feb 22. This might imply that other countries had set a higher detection priority to people traveling from Iran since Feb 19. For travelers from other places, the time intervals are summarized as follows. Wuhan, China has a median of 4.10 days and a 95\% credible interval of (1.01,10.10) days; United States has a median of 6.01 days and a 95\% credible interval of (1.69,13.90) days; Egypt has a median of 9.28 days and a 95\% credible interval of (2.44, 21.10) days; Italy has a median of 9.64 days and a 95\% credible interval of (3.86,18.79) days.


The time to diagnosis was further stratified by destinations using box plots in each sub-figure, and the destinations were arranged in an increasing order of the median time interval from top to bottom. We only presented results for destinations that had reported the arrival dates within the study period. If a country had imported cases within the study period but did not release the arrival date of any case, then its median time interval would be the overall median across other destinations that had reported the arrival dates indicated by the pink dashed line. For China, Iran and Italy, most of the destinations are their neighboring countries. Before Jan 23, 2020, seven destinations detected COVID-19 cases among travelers from Wuhan. They were South Korea, Taiwan, Singapore, Thailand, United States, Japan, and Vietnam. By Aug 14, 2020, the numbers of reported COVID-19 related deaths were 7 in Taiwan, 21 in Vietnam, 27 in Singapore, 58 in Thailand, 305 in South Korea and 1,073 in Japan \citep{JHUdata}. Those six places are among the strongest performers in the COVID-19 pandemic so far. The early detection of emerging infectious diseases among travelers did not only ring the alarm for the travel origin, but also allowed the destinations better prepare for the potential pandemic.

\subsection{Exponential growth rate, doubling time, and the probability of severe COVID-19 cases exceeding the hospital capacity}
\label{sec:beta1}


\begin{figure}
\begin{subfigure}[b]{\textwidth}
  \centering
  \caption{$\hat{\beta}_1$ (left) and $\hat{T}_d$ (right) in Wuhan, China between Jan 13 and Jan 23}
  \includegraphics[width=0.49\linewidth]{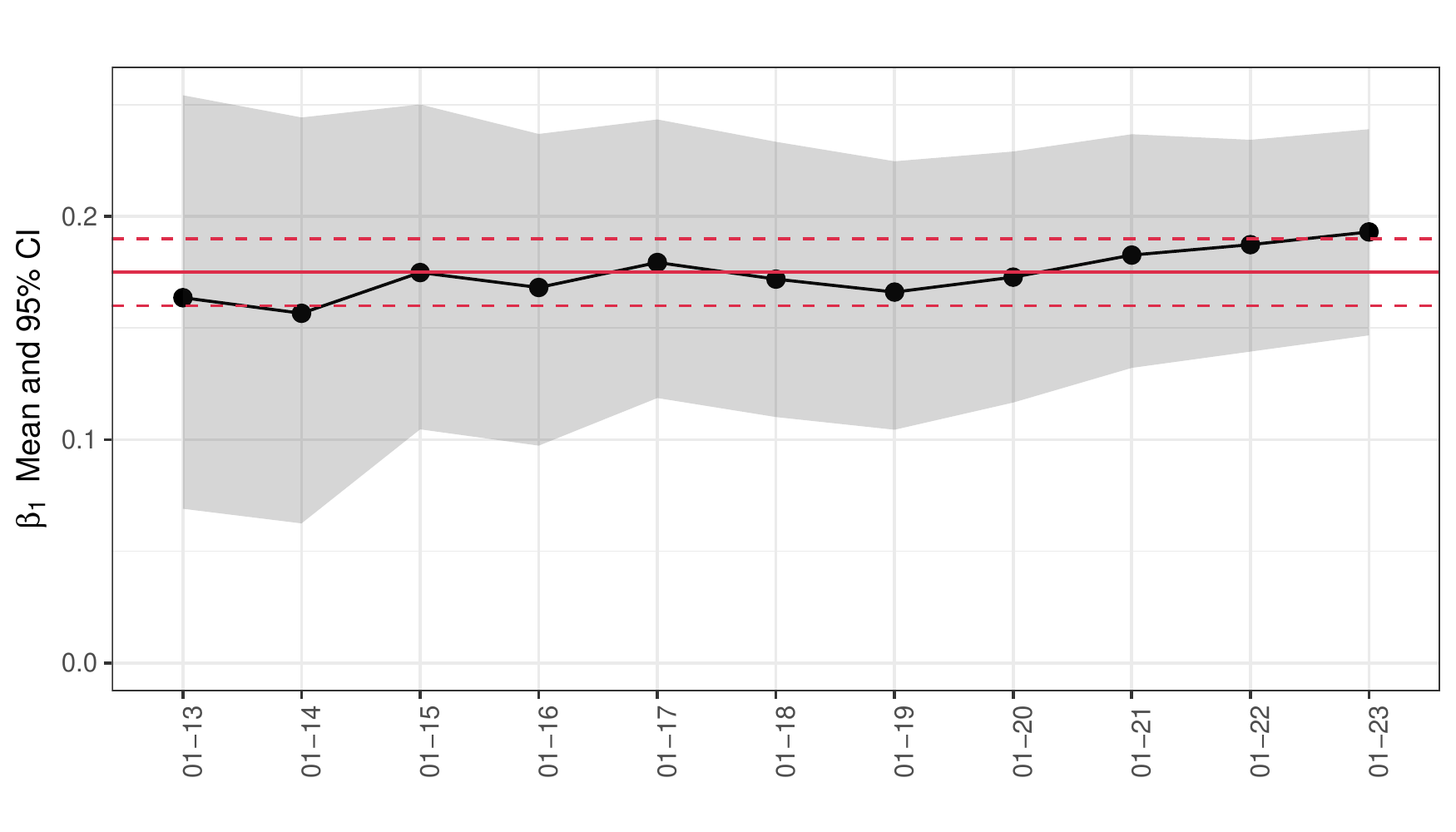}
\hfill
  \includegraphics[width=0.49\linewidth]{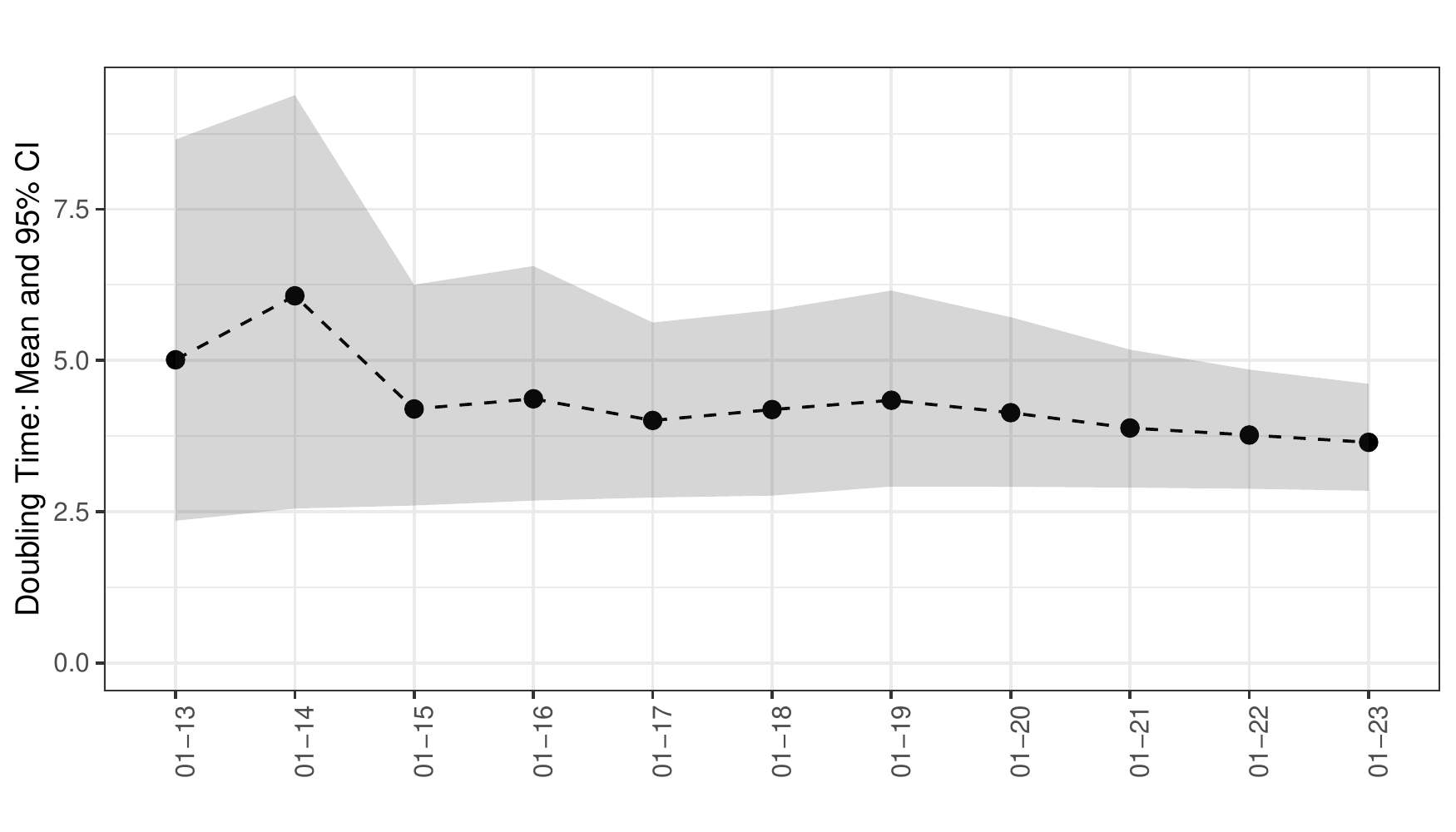}  
\end{subfigure}

\begin{subfigure}[b]{\textwidth}
  \centering
  \caption{$\hat{\beta}_1$ (left) and $\hat{T}_d$ (right) in United States between Mar 6 and Mar 13}
  \includegraphics[width=0.49\linewidth]{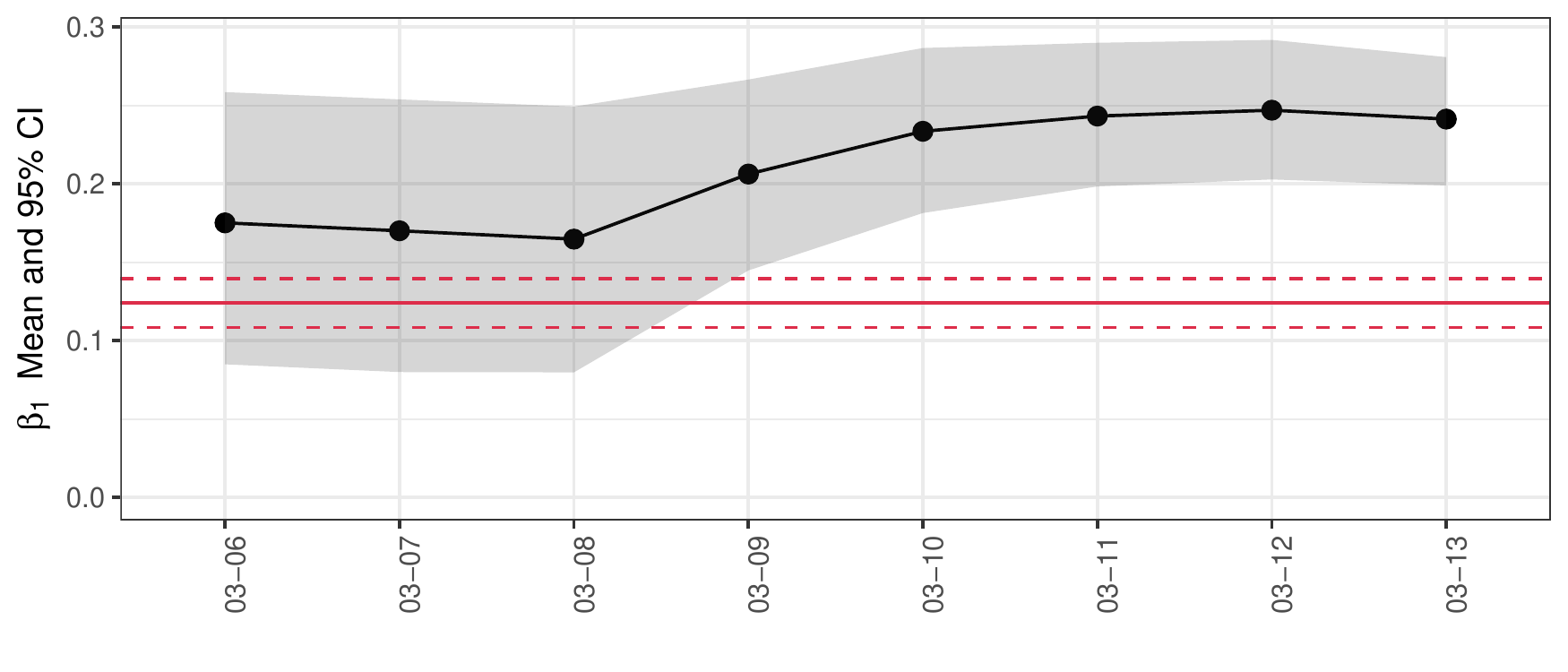}
\hfill
  \includegraphics[width=0.49\linewidth]{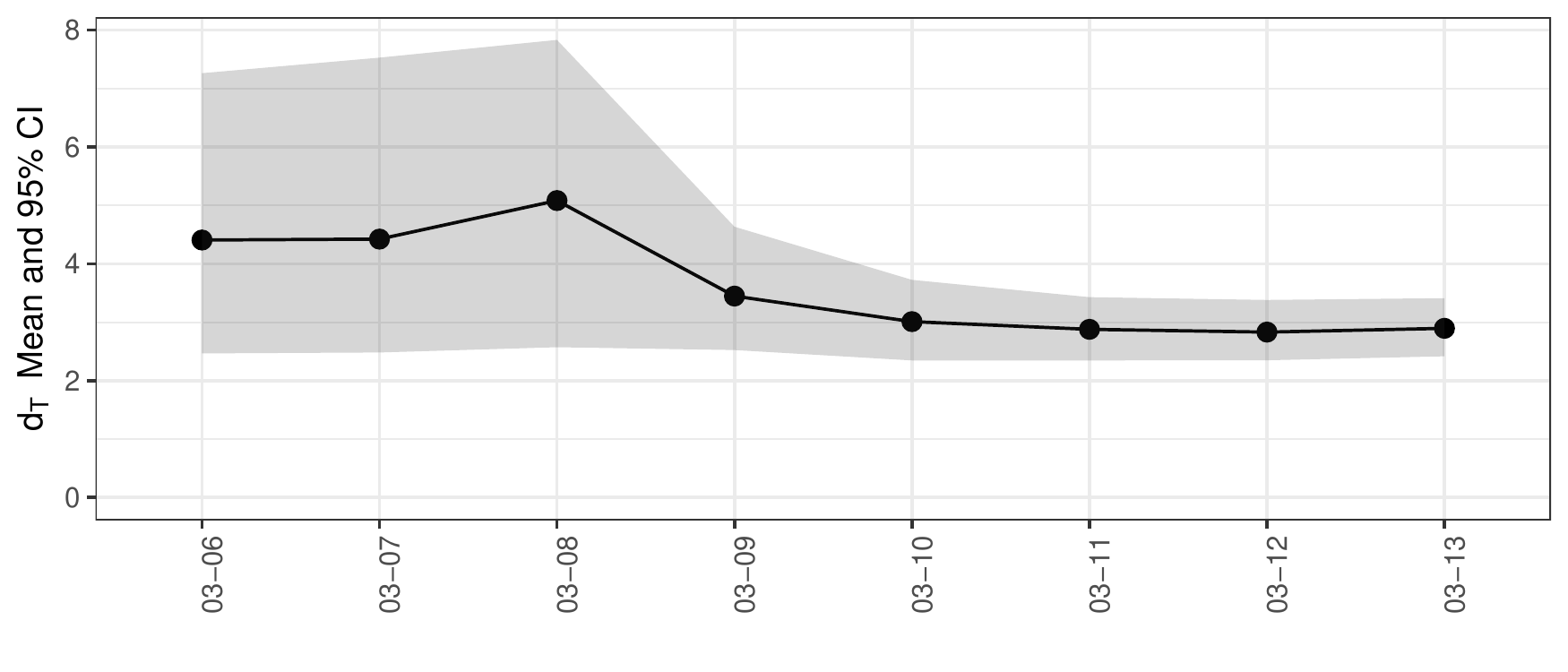}  
\end{subfigure}

\begin{subfigure}[b]{\textwidth}
  \centering
  \caption{$\hat{\beta}_1$ (left) and $\hat{T}_d$ (right) in Italy between Feb 25 and Feb 28}
  \includegraphics[width=0.49\linewidth]{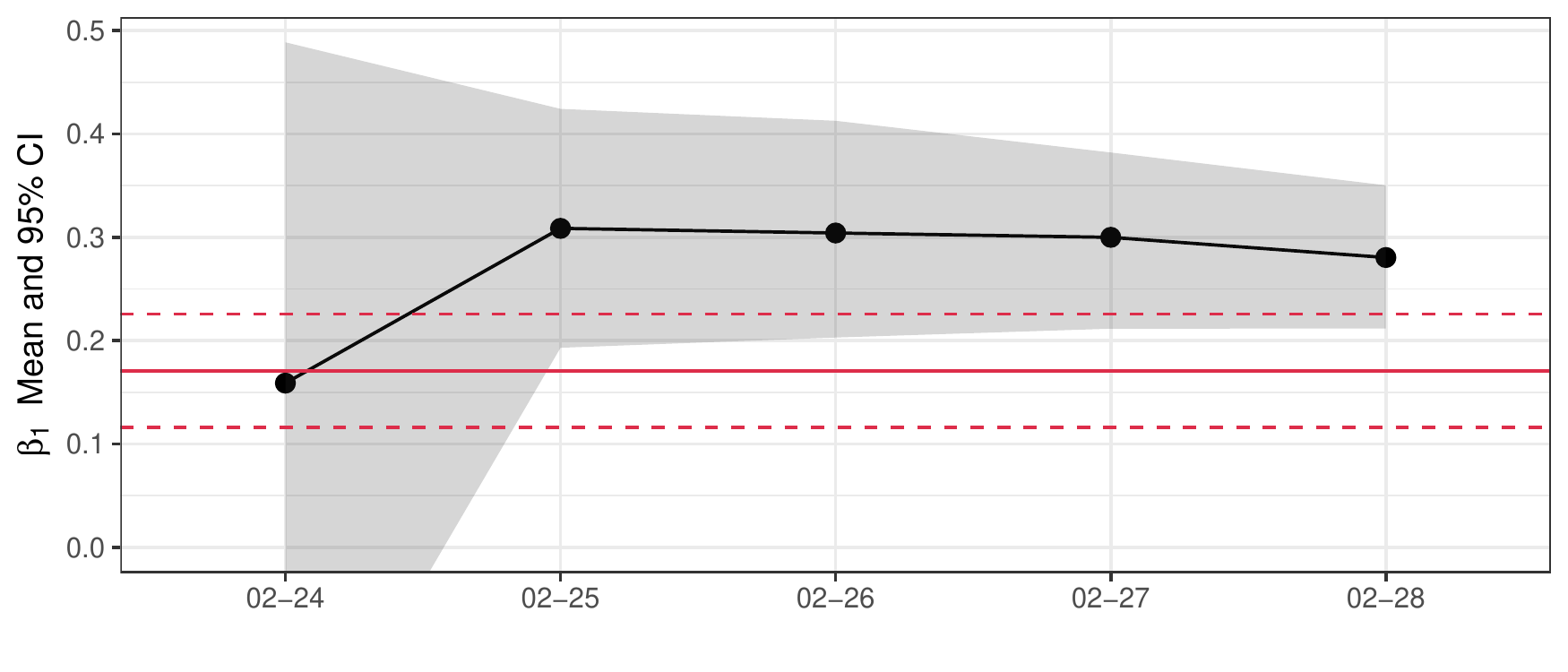}
\hfill
  \includegraphics[width=0.49\linewidth]{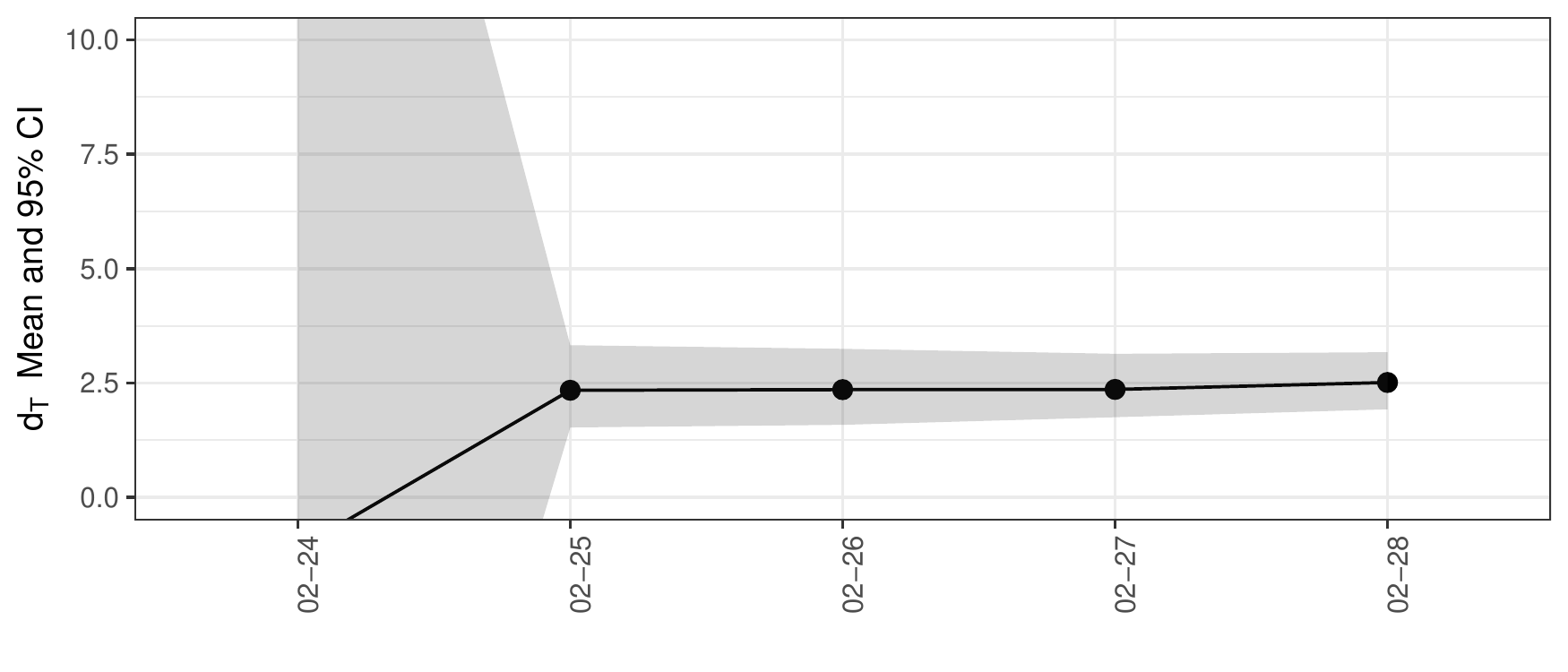}  
\end{subfigure}

\begin{subfigure}[b]{\textwidth}
  \centering
  \caption{$\hat{\beta}_1$ (left) and $\hat{T}_d$ (right) in Iran between Feb 20 and Feb 27}
  \includegraphics[width=0.49\linewidth]{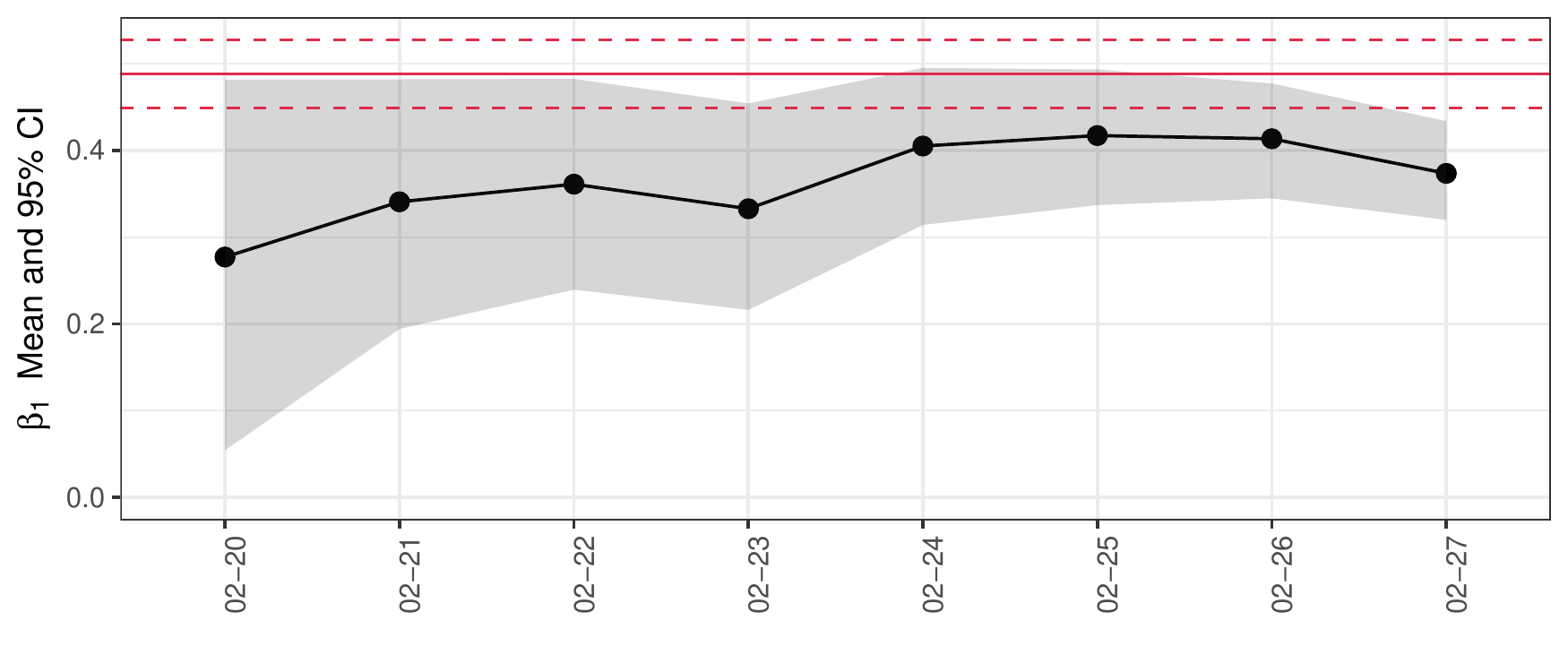}
\hfill
  \includegraphics[width=0.49\linewidth]{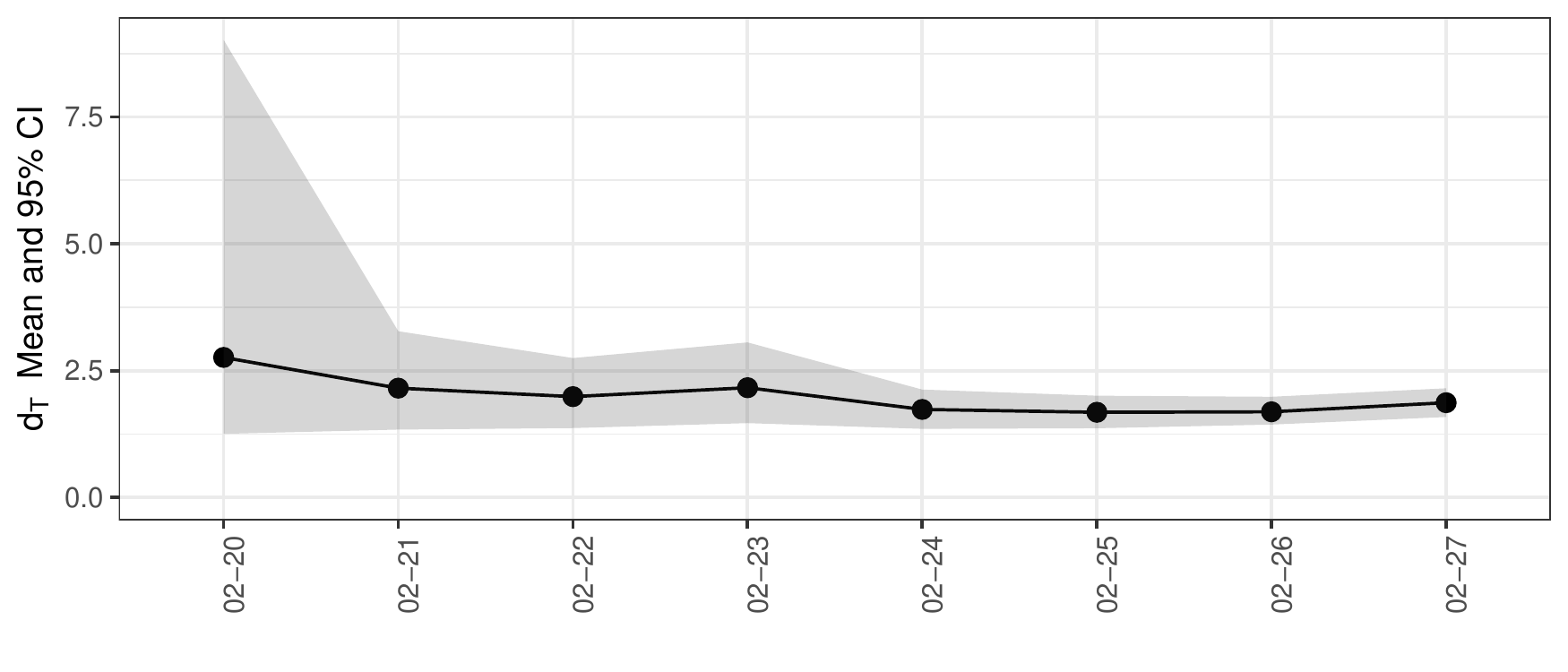}  
\end{subfigure}

\begin{subfigure}[b]{\textwidth}
  \centering
  \caption{$\hat{\beta}_1$ (left) and $\hat{T}_d$ (right) in Egypt between Feb 28 and Mar 6}
  \includegraphics[width=0.49\linewidth]{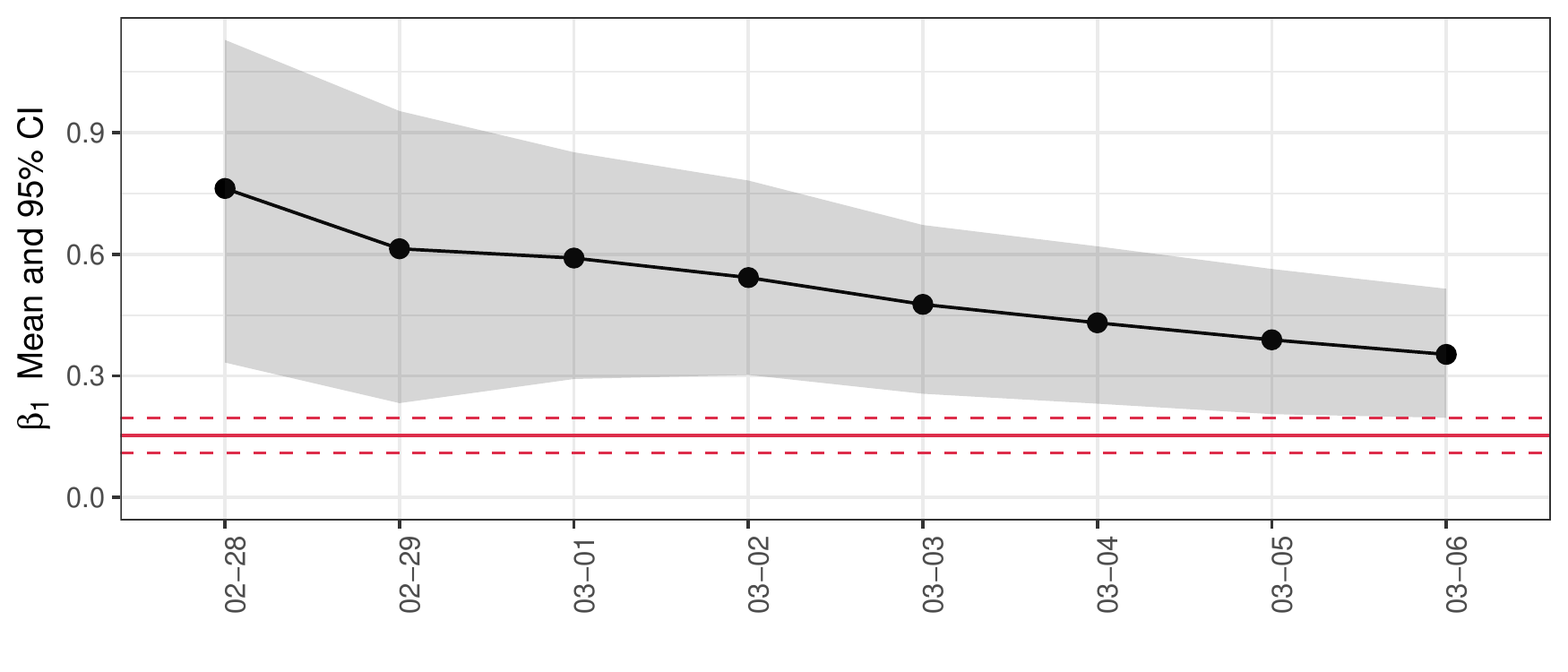}
\hfill
  \includegraphics[width=0.49\linewidth]{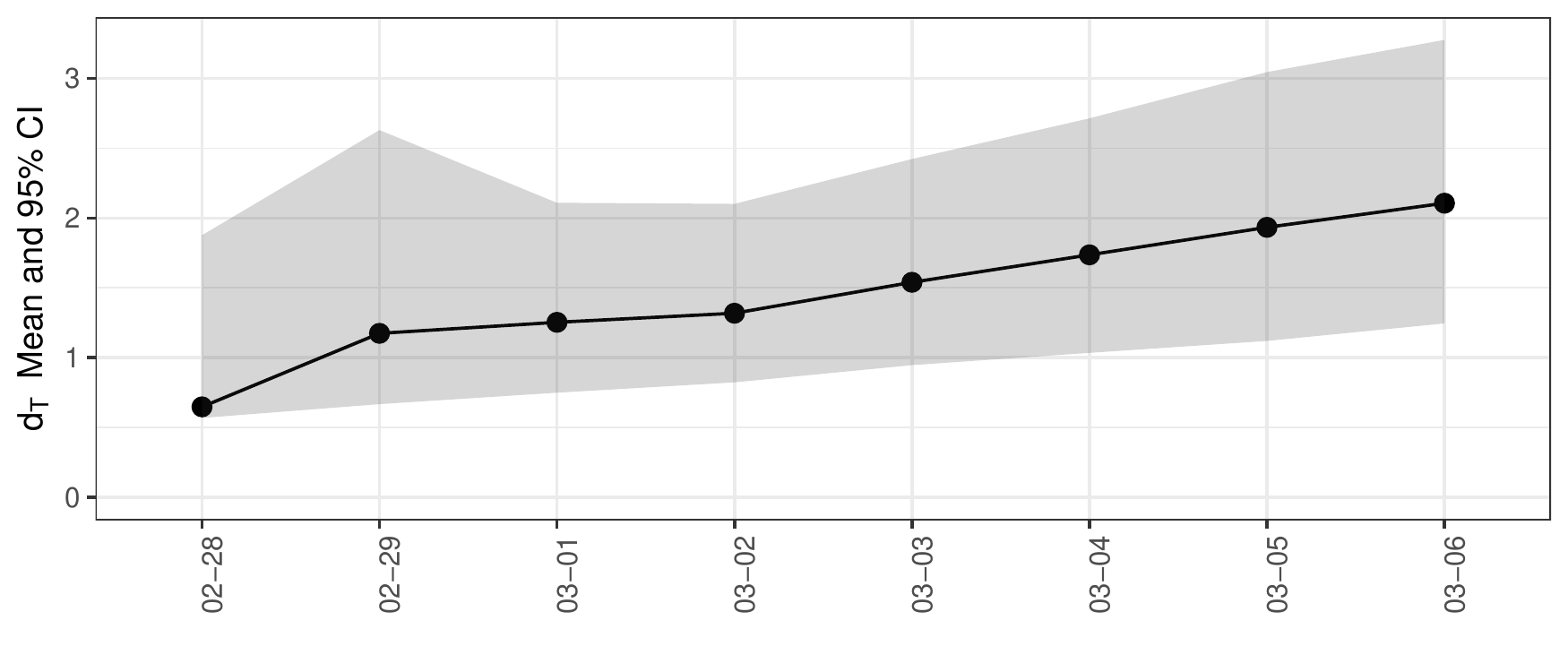}  
\end{subfigure}
\caption{Posterior mean and 95\% credible interval of exponential growth rate $\beta_1$ at early COVID-19 outbreak stage in (a) Wuhan, China, (b) United States, (c) Italy, (d) Iran, (e) Egypt. The black dots are the posterior mean and the grey bands are the 95\% confidence credible intervals estimated from traveler data. The red horizontal lines indicate the mean (solid) and 95\% confidence interval (dashed) estimated from domestic data.}
\label{fig:trend}
\end{figure}

For each origin, we obtained estimates of the exponential growth rate, $\beta_1$, and the doubling time, $T_d$, by fitting the model described in Section 2.3 to the traveler case reports up to different dates. Figure \ref{fig:trend} visualizes how the mean estimates and 95\% uncertainty bounds of $\beta_1$ and $T_d$ changed as more confirmed cases became available over time. 



The sequential updates of $\hat{\beta}_1$ and $\hat{T}_d$ became more stable over time. For all travel origins, within three days after the initial exported case report, the estimated lower bound of the exponential growth rate was above 0.1, which corresponded to the doubling time being significantly less than one week. Table \ref{tab:dates} summarizes the dates on which the estimated exponential growth rate, $\beta_1$, was significantly greater than $0.1$. We found less than ten exported case reports would be needed and the detection dates were within three days after the first exported case report for all countries.

\begin{table}[!ht]
    \caption{The dates on which the estimated exponential growth rate, $\beta_1$, was significantly greater than $0.1$. For each date, we also provide the number of days after the first exported case confirmation and the cumulative number of exported case reports.}
    \label{tab:dates}
        \centering
\begin{tabularx}{0.95\textwidth}{c *{3}{Y}}
    \toprule
    & detection date & number of days & number of cases \\
        \midrule
      Wuhan & Jan 15 & 2 days & 2 cases \\
      U.S. & Mar 9 & 3 days & 3 cases \\
      Italy & Feb 25 & 1 day & 9 cases \\
      Iran & Feb 21 & 1 day & 2 cases \\
      Egypt & Feb 28 & 0 day & 1 case \\
\bottomrule
\end{tabularx}
\end{table}

The estimated exponential growth rate in Wuhan, Iran, Italy and United States showed an overall increasing trend at the beginning of the outbreak after the confirmation of the first exported case. The decreasing trend of $\hat{\beta}_1$ in Egypt was because we assumed the epidemic in Egypt started on Feb 14, 2020 with only one infected individual on that date. A high level of growth rate would be needed to observe many traveler cases between March 1, 2020 and March 6, 2020. Since the travel date of the first exported case from Italy (confirmed on Feb 24, 2020) was unknown, $\hat{\beta}_1$ and $\hat{d}_T$ had a larger uncertainty when only observing the first exported case. 

To compare the estimates from travelers data with the ones from domestic data, we fitted exponential growth curves to the domestic daily records using \cite{Pan2020association} for Wuhan (China) and the COVID-19 Data Repository \citep{JHUdata} for United States, Italy, Iran and Egypt. To maximize the sample size, we chose the end date of the domestic data to be 5 days (the mean incubation period) after the country announced national emergency or implemented lock-down order. The mean estimates and confidence intervals of the growth rates were 0.175 (0.160, 0.190) for Wuhan (China), 0.124 (0.108, 0.140) for United States, 0.171 (0.116, 0.226) for Italy, 0.488 (0.449, 0.527) for Iran, 0.153 (0.110, 0.196) for Egypt. We indicated these estimates and intervals with horizontal red lines in Figure \ref{fig:trend}. 
The domestic data mean estimates for United States, Italy and Egypt were significantly lower than the traveler data based estimates. Some possible reasons for the discrepancy include: (a) their initial numbers of cases were larger (see sensitivity analysis in Section \ref{sec:sensitivity-initial}); (b) the number of infected increased faster than the testing capacity in the early period; (c) the number of travelers had declined within the study period; and (d) the travelers did not represent the general population thus could bear a higher or lower infection rate. (a) and (b) could lead to underestimated growth rate using domestic data. (c) could lead to overestimated growth rate using traveler data. (d) could lead to either direction of biases.

\begin{figure}[!ht]
\centering
\includegraphics[width=0.7\linewidth]{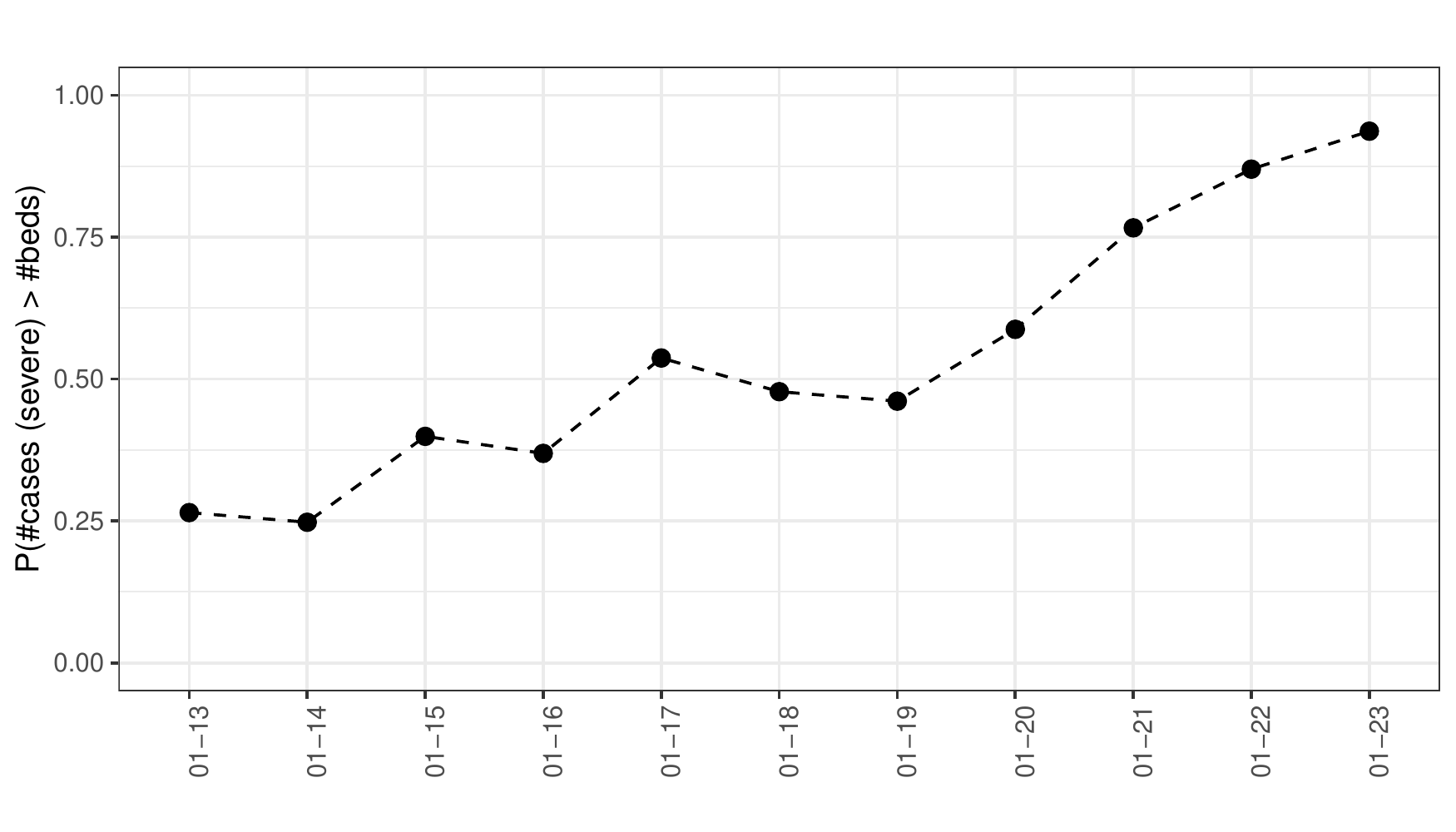}  
\caption{Posterior probability of the number of severe cases exceeds the number of available hospital beds in Wuhan China during Jan 13, 2020 - Jan 23, 2020.}
\label{fig:prob-wuhan}
\end{figure}

In Figure \ref{fig:prob-wuhan}, we plotted the probability of severe COVID-19 cases exceeding the number of available beds in Wuhan. With the first exported case confirmation on Jan 13, the probability was at a relatively low level around 26\%. It jumped to 54\% when the third exported case was confirmed on Jan 17, and kept increasing as new case reports became available over time. The probability exceeded 90\% on Jan 23, indicating that hospitals would soon be overwhelmed with COVID-19 patients. 



\subsection{Sensitivity analysis to the initial number of cases}
\label{sec:sensitivity-initial}

In the previous analysis, we assumed the epidemic start date for each origin as the date when they announced their first case(s), denoted as $t_0$. In this section, we investigated how the estimate of the exponential growth rate, $\hat{\beta_1}$, would have changed if the epidemic had started earlier. An earlier starting date is equivalent to there were more cases at $t_0$ than 1 or 2. We reran the analysis with the  number of cases at $t_0$ being 10, 100, and 1000 respectively. 

\begin{figure}[!ht]
  \centering
\begin{subfigure}[b]{0.495\textwidth}
    \subcaption{$\hat{\beta}_1$ in Wuhan between Jan 13 and Jan 23}
    \includegraphics[width=\linewidth]{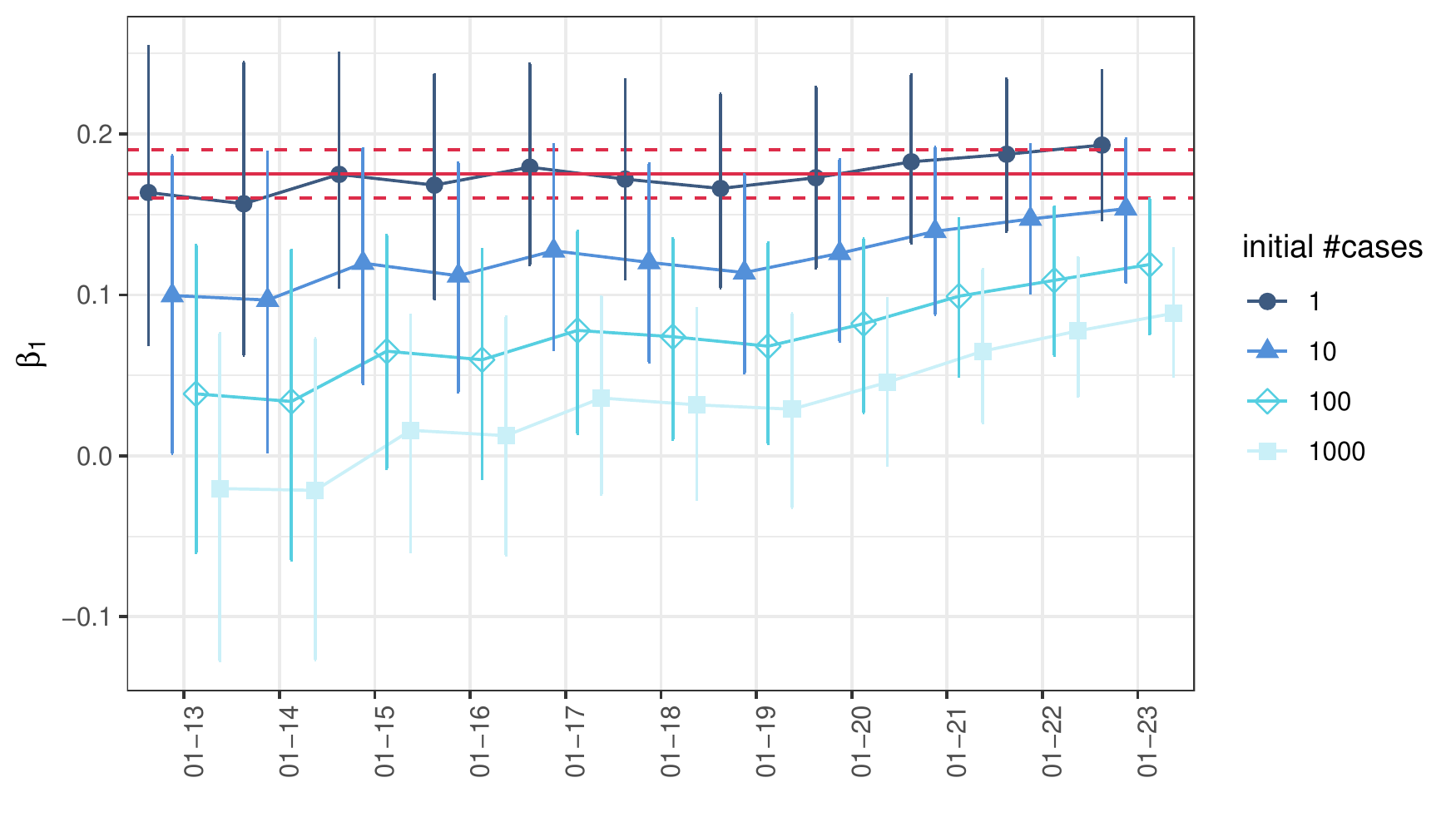}
\end{subfigure}
\hfill
\begin{subfigure}[b]{0.495\textwidth}
    \subcaption{$\hat{\beta}_1$ in United States between Mar 6 and Mar 13}
    \includegraphics[width=\linewidth]{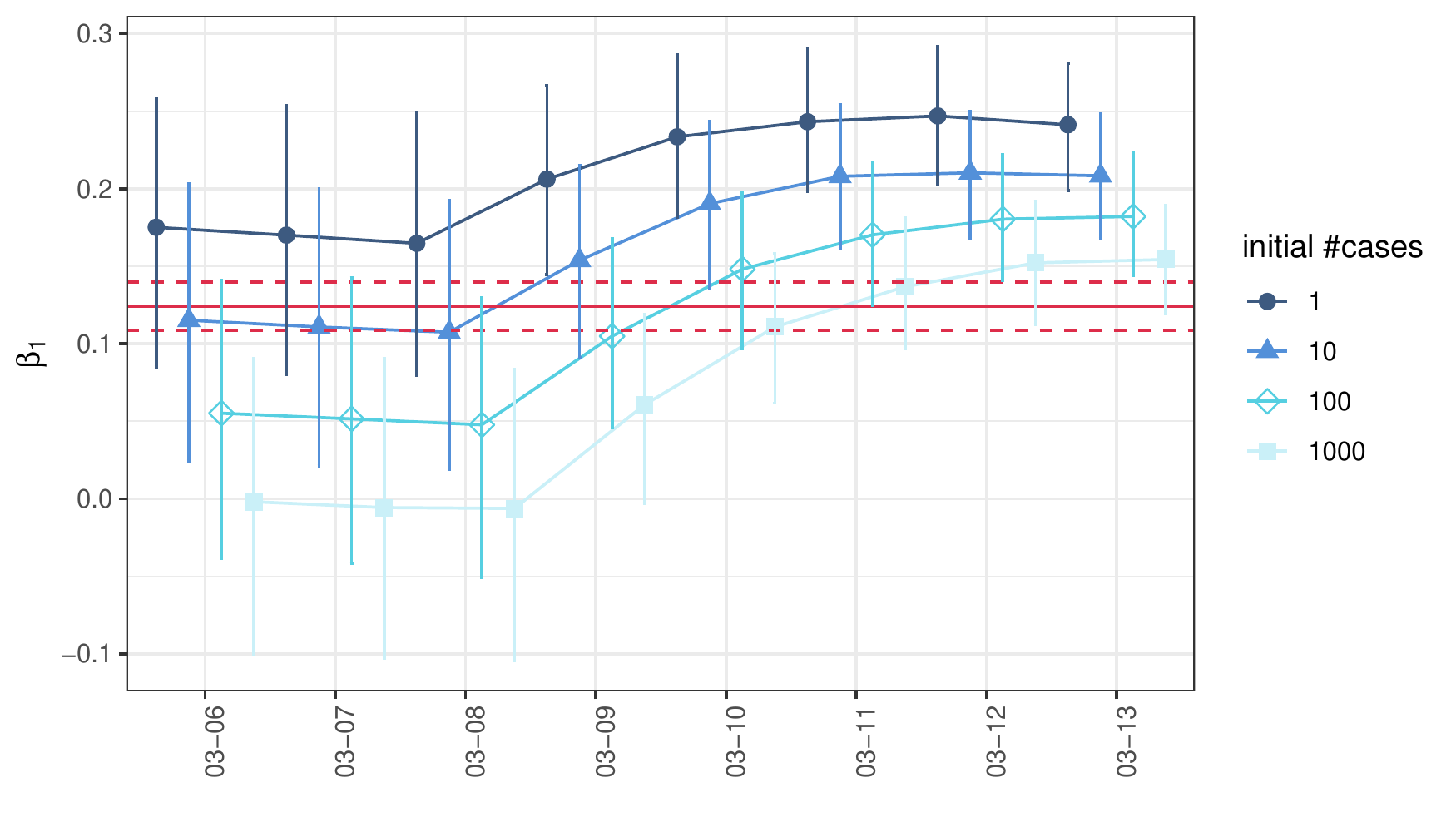}
\end{subfigure}
\bigskip
\begin{subfigure}[b]{0.495\textwidth}
    \subcaption{$\hat{\beta}_1$ (left) in Italy between Feb 24 and Feb 28}
    \includegraphics[width=\linewidth]{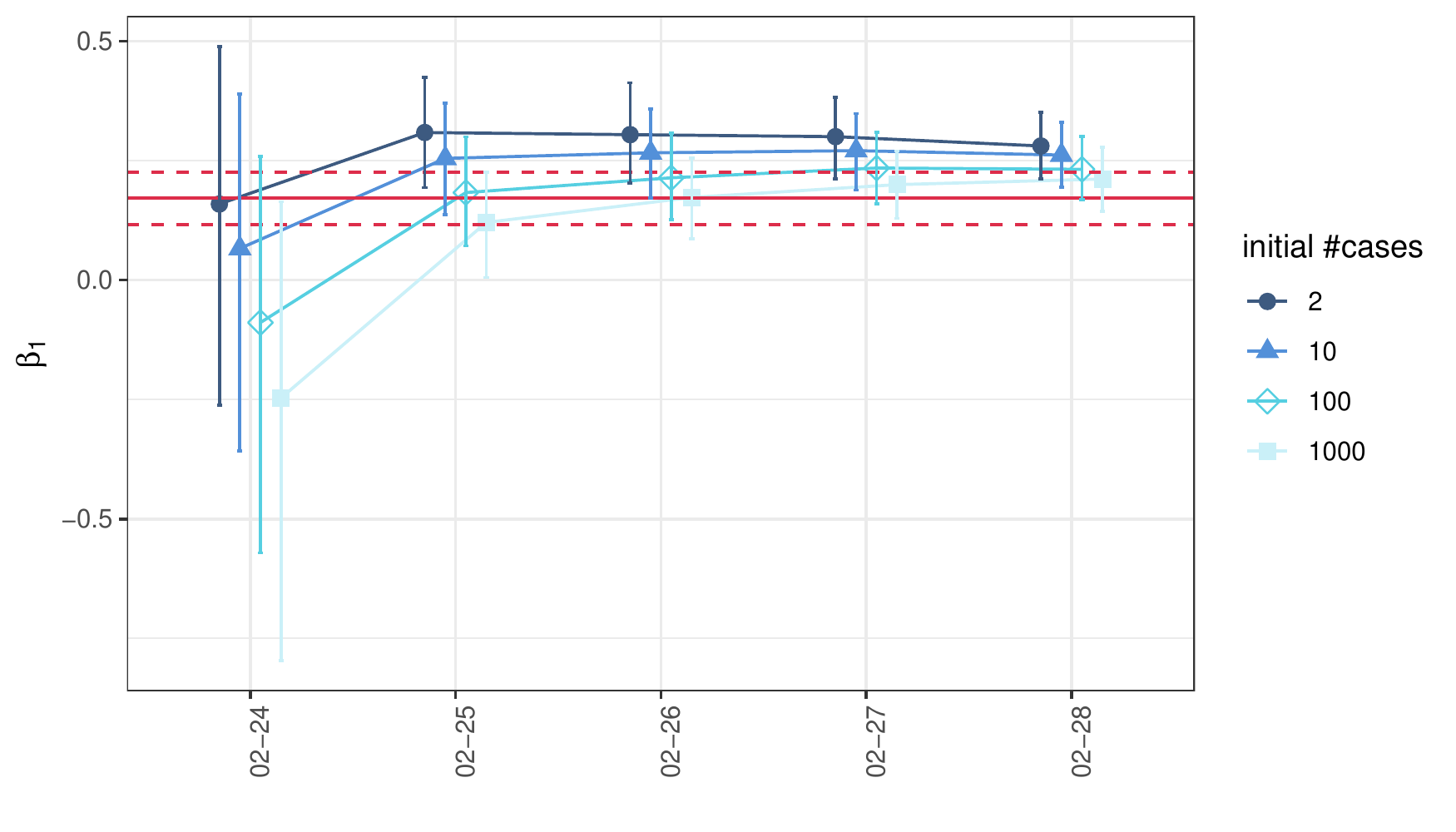}
\end{subfigure}
\hfill
\begin{subfigure}[b]{0.495\textwidth}
    \subcaption{$\hat{\beta}_1$ in Iran between Feb 20 and Feb 27}
    \includegraphics[width=\linewidth]{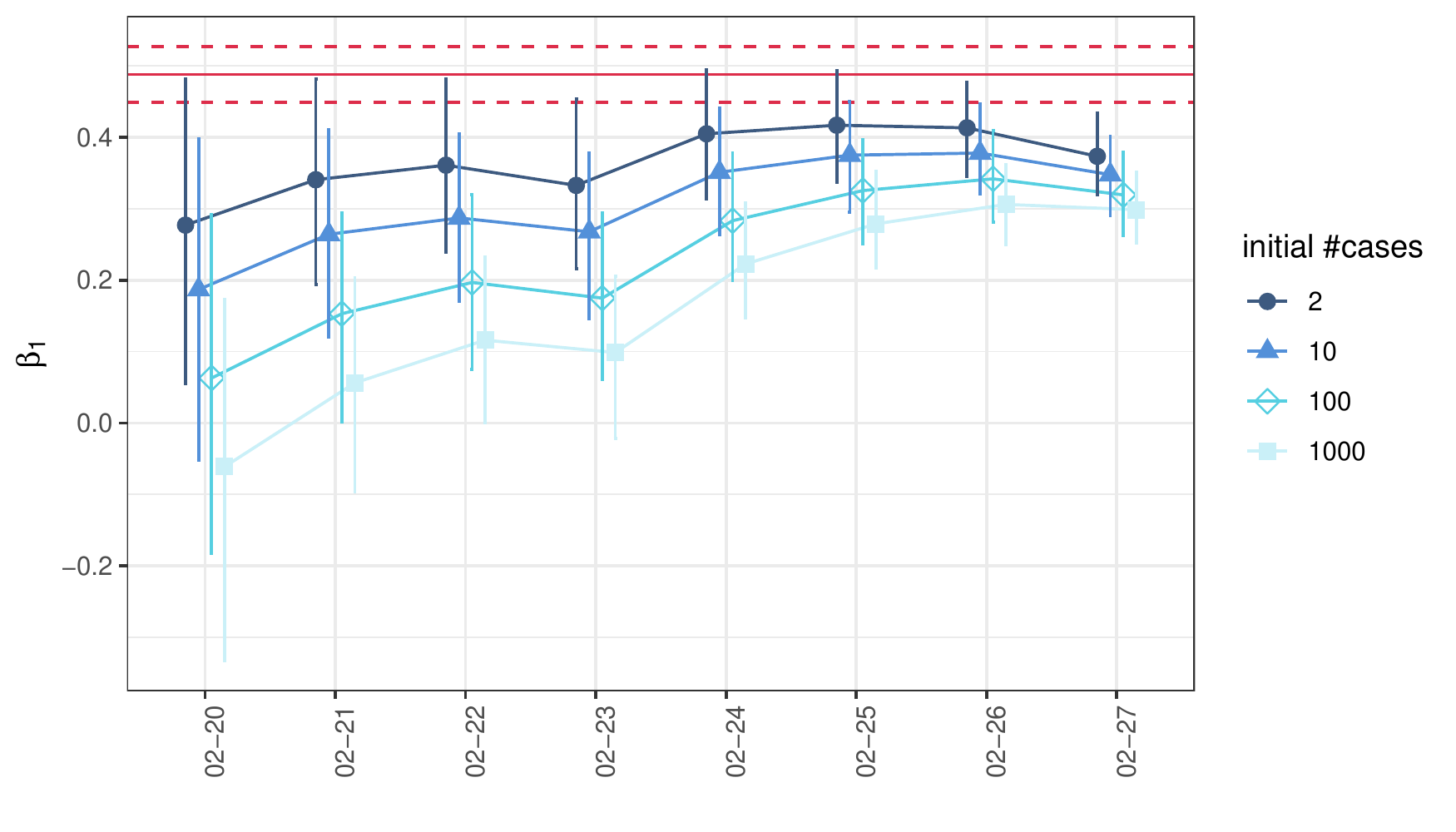}
\end{subfigure}
\bigskip
\begin{subfigure}[b]{0.495\textwidth}
    \subcaption{$\hat{\beta}_1$ in Egypt between Feb 28 and Mar 6}
    \includegraphics[width=\linewidth]{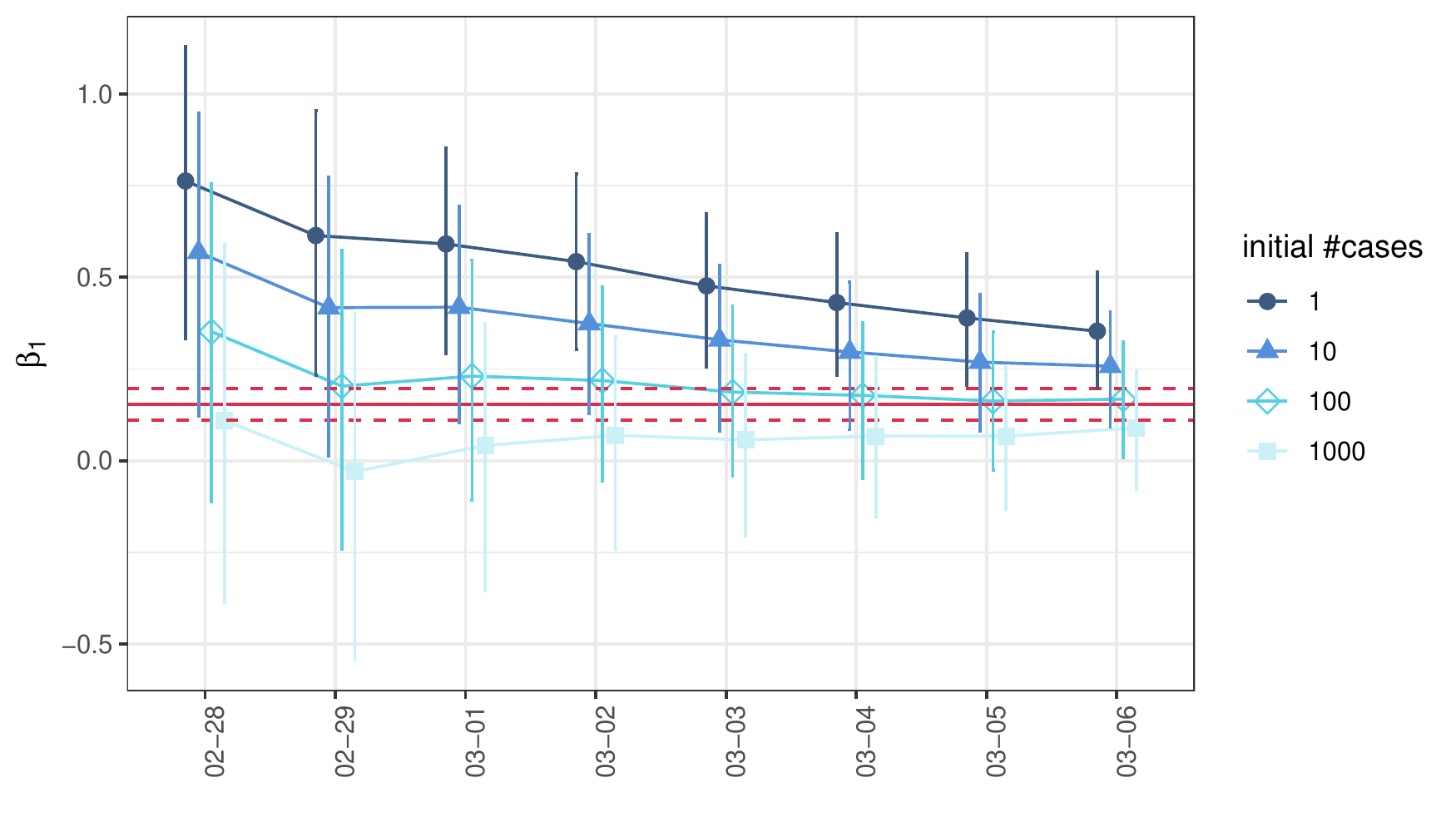}
\end{subfigure}
\hfill
\begin{subfigure}[b]{0.495\textwidth}
    \subcaption{Probability of the number of severe case patients exceeds the hospital capacity in Wuhan, China}
    \includegraphics[width=\linewidth]{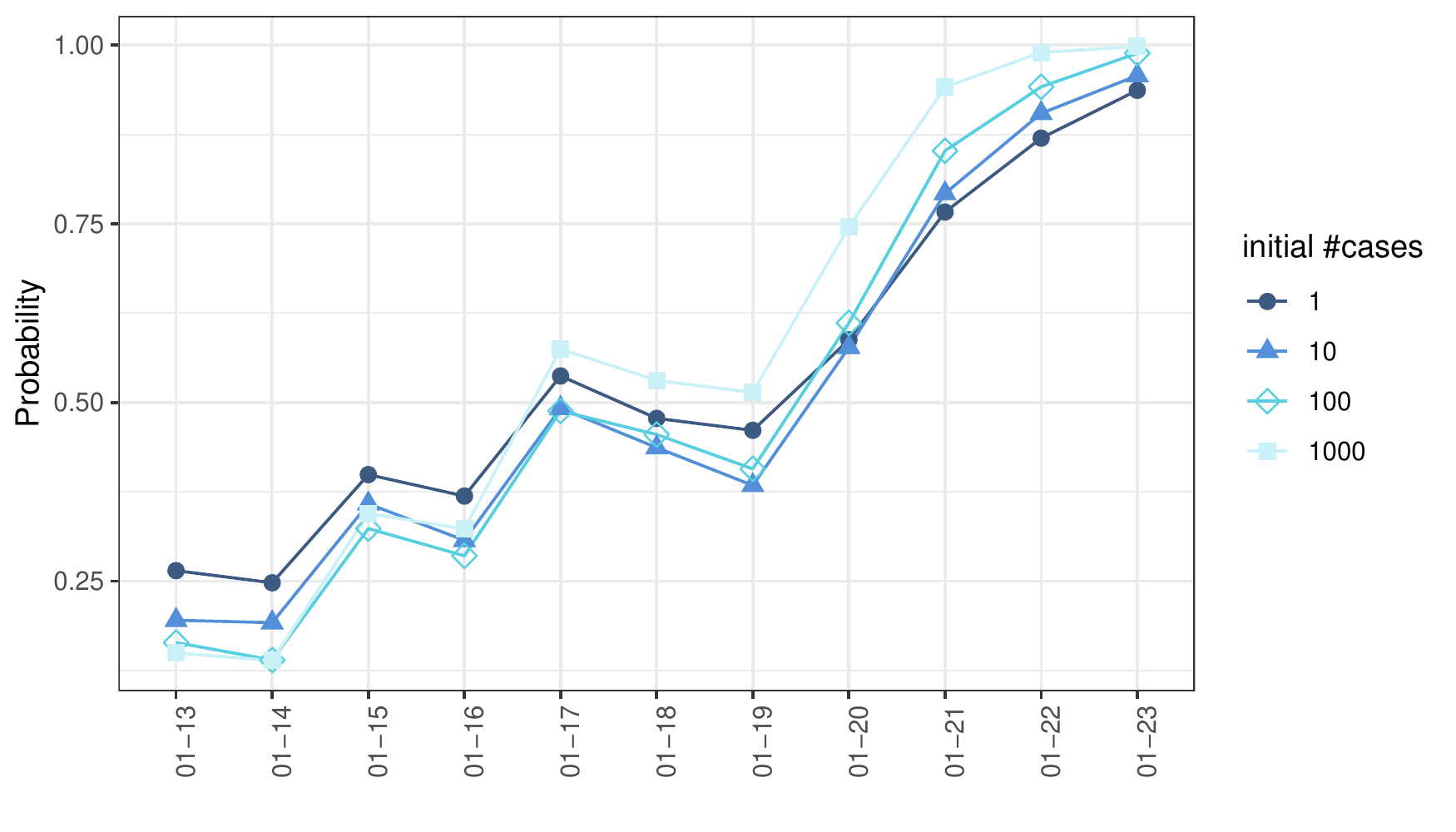}
\end{subfigure}
\caption{The estimated exponential growth rate, $\hat{\beta}_1$, under different initial number of cases in (a) Wuhan, China, (b) United States, (c) Italy, (d) Iran, (e) Egypt. The red horizontal lines indicate the mean (solid) and 95\% credible interval (dashed) estimated from domestic data. (g) the estimated probability of the number of severe case patients exceeds the hospital capacity under different initial number of cases in Wuhan, China. The dots are the posterior mean and error bars are the 95\% confidence credible intervals. Different color represents different initial number of cases.}
\label{fig:initial}
\end{figure}

Figure \ref{fig:initial} presents the estimated exponential growth rates using traveler data and different initial number of cases, as well as the exponential growth rates estimated from domestic data (red horizontal lines). The black lines indicated the original model fits assuming that the number of initial cases was either 1 or 2 at $t_0$. A larger number of initial cases corresponded to a lower estimate of the exponential growth rate. For Wuhan and Italy (panels (a) and (c) in Figure \ref{fig:initial}), the current assumed initial case number (epidemic start date) seemed to be reasonable. For Iran (panel (d)), the first government announcement in Iran regarding to COVID-19 was two deaths on Feb 19, 2020 \citep{world2020coronavirus}. We assumed that the start date was 20 days before, based on that the first COVID-19 related death in Wuhan was on Jan 9, 2020 with symptom onset on Dec 20, 2019. This appeared to be a reasonable assumption given the traveler data estimates (black) crossing the domestic data estimated interval (red) in the end. For United States, Italy and Egypt (panels (b), (c) and (e))), their 95\% credible intervals at the end date of study period were above the domestic data estimates which better matched with ``100 cases" scenario or ``1000 cases" scenario.

There were two reasons that we might set the initial number of cases at a higher level: one was the under-reporting of initial cases; and the other was the epidemic being limited to a sub-national area where the ratio between the traveler size and the local population size was much higher than the national average. Note that we had used the national population as the reference to calculate the initial COVID-19 prevalence rates in United States, Italy, Iran and Egypt. If the outbreak was limited within a sub-national area, one should adjust both the reference population size and the traveler size from the national level to the sub-national level. This would not affect the estimate of the growth rate if the ratio between the traveler size and reference population size at the sub-national level is equal to that ratio at the national level. However, for areas such as tourism sites, the ratio between traveler size and the reference population size could be much higher than the national average. For instance, \citep{Negida2020estimation} commented that ``Egypt is an agricultural country. Most tourism destinations are in special locations far away from residential places and have low population densities." If the Egypt outbreak was indeed only at the tourism destinations where the reference population size was small but the international traveler volume was large, then $\hat{\beta}_1$ for the tourism destinations in Egypt should be similar to ``100 cases" or ``1000 cases" scenario in Figure \ref{fig:initial} (e), which did not have the surprisingly high estimates in the first couple of days, and better matched with the estimates from the domestic daily records. Unfortunately, most traveler case reports only provided the countries being visited but not the sub-national areas.


We also noticed that the prevalence $\rho_t$ in the origin was less sensitive to the assumption of the start date. Therefore, the total number of cases and severe cases were similar in different scenarios. As seen in Figure \ref{fig:initial} (f), the probabilities of the number of severe cases exceeding the number of available hospital beds in Wuhan China were similar.

Finally, we found that the differences of $\hat{\beta}_1$s among different scenarios of initial cases became smaller as more traveller case reports became available over time. The effect of initial number of cases will be addressed quantitatively in Section \ref{sec:VOI}.

\subsection{Impact of daily traveler case reports and travel dates}
\label{sec:VOI}

As described in the Methods section, we conducted a series of analysis on the  Wuhan outbound traveler case report data to evaluate the influence of different information on each day's estimate. Every time a new case report was observed, we computed the value of information of all of the previous case reports on making the current day's decision (estimation). The case reports among Wuhan travelers were available on Jan 13 (one case), Jan 15 (one case), Jan 17 (one case), Jan 20 (one case), Jan 21 (two cases), Jan 22 (two cases), and Jan 23 (three cases). Figure \ref{fig:voi_all} presents the value of information (VOI) analysis results. Figure \ref{fig:voi_all} was organized by decision (estimation) date. In each sub-figure, we calculated the VOI of all of the case reports up to this date on estimating the current $\beta_1$, and displayed them using the blue line. We also calculated the VOI of knowing each arrival date on estimating the current $\beta_1$, and displayed them using the red line. We had the following findings:

\begin{figure}[!ht]
  \centering
\includegraphics[width=\linewidth]{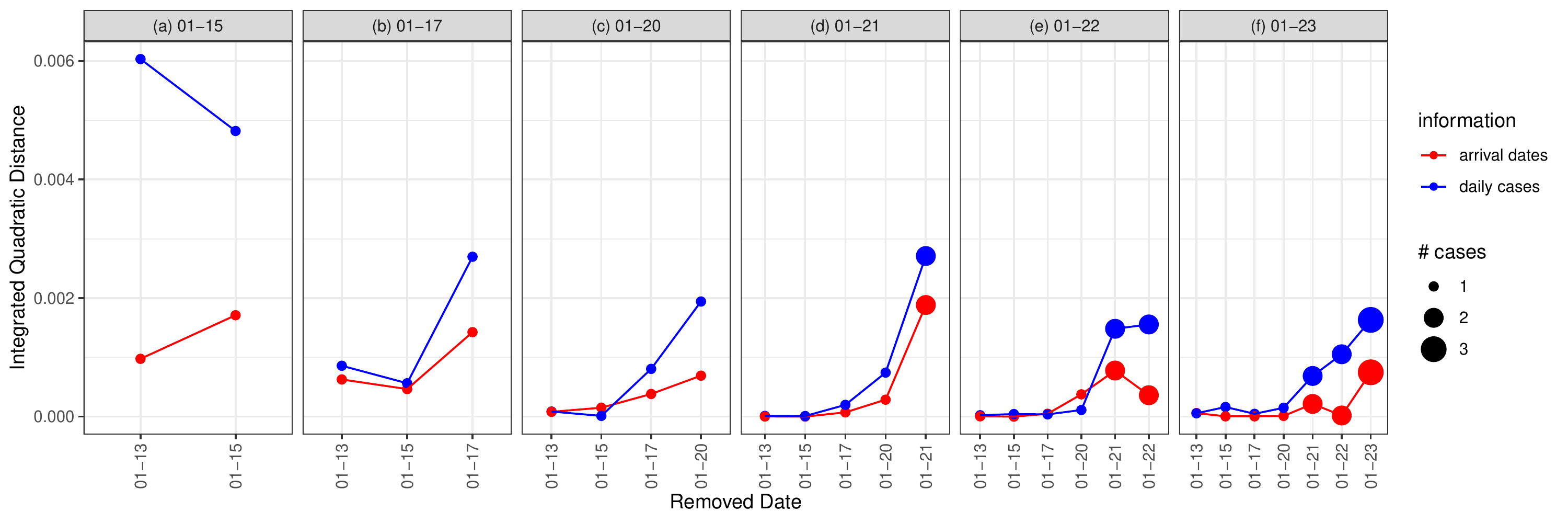}
\caption{The value of information (VOI) of daily case and arrival date reports during Jan 13 - Jan 23. The blue lines are the VOIs of daily case reports. The red lines are the VOIs of daily arrival date reports. Each sub-figure corresponds to estimating $\beta_1$ on a different date $t$, shown as the sub-title.}
\label{fig:voi_all}
\end{figure}

\begin{enumerate}[wide, labelwidth=!, labelindent=0pt]
    
    \item For the days with the same number of cases, we observed that the more recently confirmed cases had higher VOI than the earlier cases, except that the case on 01-13 had a higher VOI than the case on 01-15. This was because for all the other cases, the earlier the arrival dates, the earlier the confirmed date (for the date with multiple confirmations we looked at the average arrival date). However, the arrival date of the 01-15 case was earlier than the 01-13 case. Therefore we concluded that the more recent arrival cases had higher VOIs.
    
    \item From sub-figures (d), (e) and (f), we observed that the days with a larger number of case confirmations had a larger VOI. 
    
    \item Each case report's VOI decreased over time. This was because first each case became less recent as we moved along the decision date, and second as more case reports became available, the expected loss of removing one case report got smaller. 
    
    \item In terms of knowing the arrival dates, in general, it was also true that the more recent arrival dates had higher VOIs. In addition, the more cases in each report, the larger its VOI was. Note that one of the two cases on 01-22 had missing arrival date, and thus the VOI of 01-22 arrival dates measured the effect of removing one arrival date.
    
    \item Comparing the VOI of the case reports and the VOI of the arrival dates, we found that in general the case reports had higher VOIs than the arrival dates and the gap was larger for more recent dates. Note that the arrival date of the 01-15 case was 01-06 (the slowest time to diagnosis) and the arrival date of the 01-20 case report was 01-19 (the fastest time to diagnosis). Their arrival date VOIs were very close to (and sometimes exceeded) the case report VOIs. It suggested the case reports would be less useful without arrival dates, especially when the time to diagnosis was much longer or shorter than the average, which made the imputation inaccurate.
\end{enumerate}


\subsection{Number of cases needed to detect the disease outbreak}
To demonstrate how many exported cases needed to detect a disease outbreak with certain statistical significance, we conducted the hypothesis test of whether the exponential growth rate exceeded 0.1. We used the initial prevalence, 0.01\%, corresponding to one initial infection in a place with one million people, and set daily average of outbound traveler size at $N=1,000$. 

\begin{figure}[!ht]
\centering
\begin{subfigure}[b]{\textwidth}
\subcaption{Total number of exported cases stratified by the number of days since the initial local infection}
\includegraphics[width=0.8\linewidth]{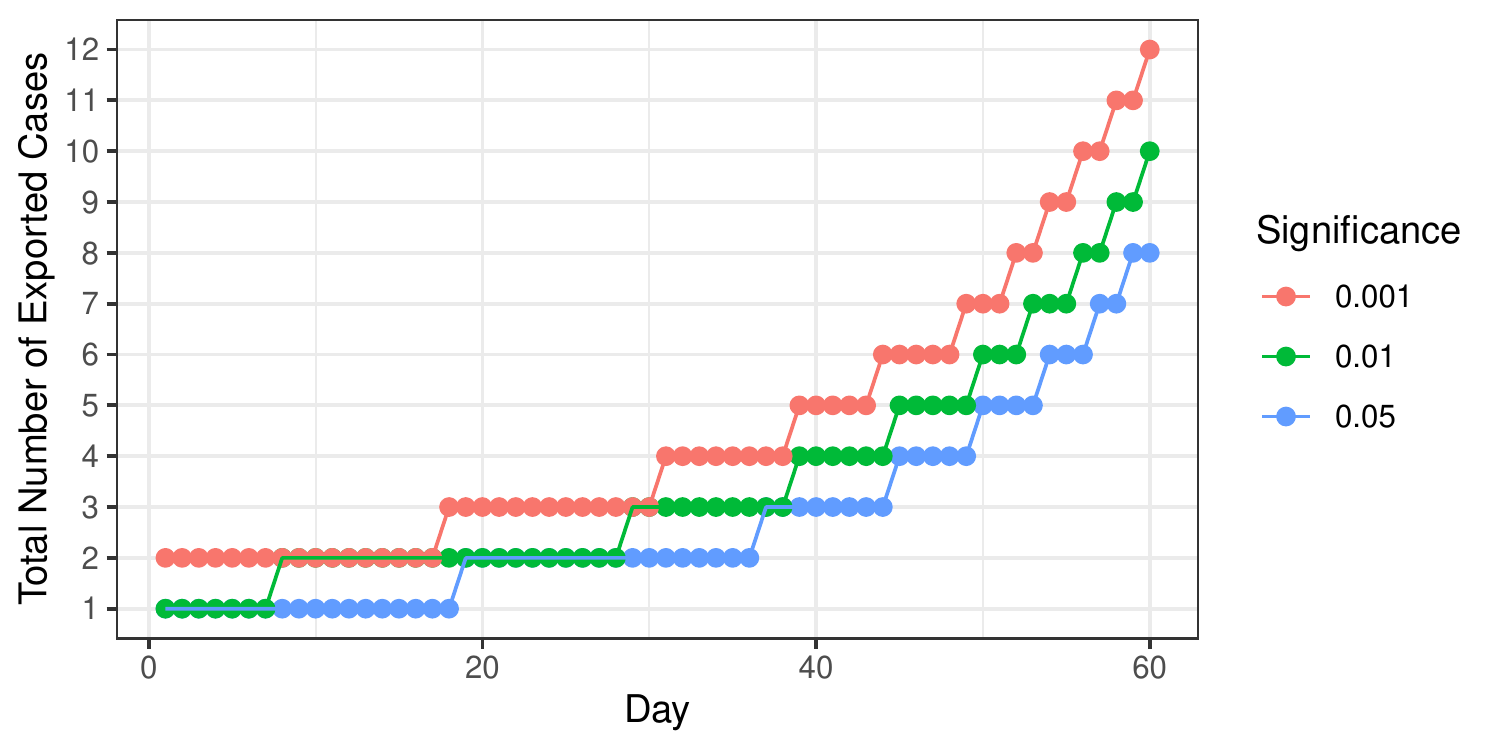}
\end{subfigure}

\begin{subfigure}[b]{\textwidth}
\subcaption{Total number of exported cases stratified by the number of days since the first exported traveler case}
\includegraphics[width=0.8\linewidth]{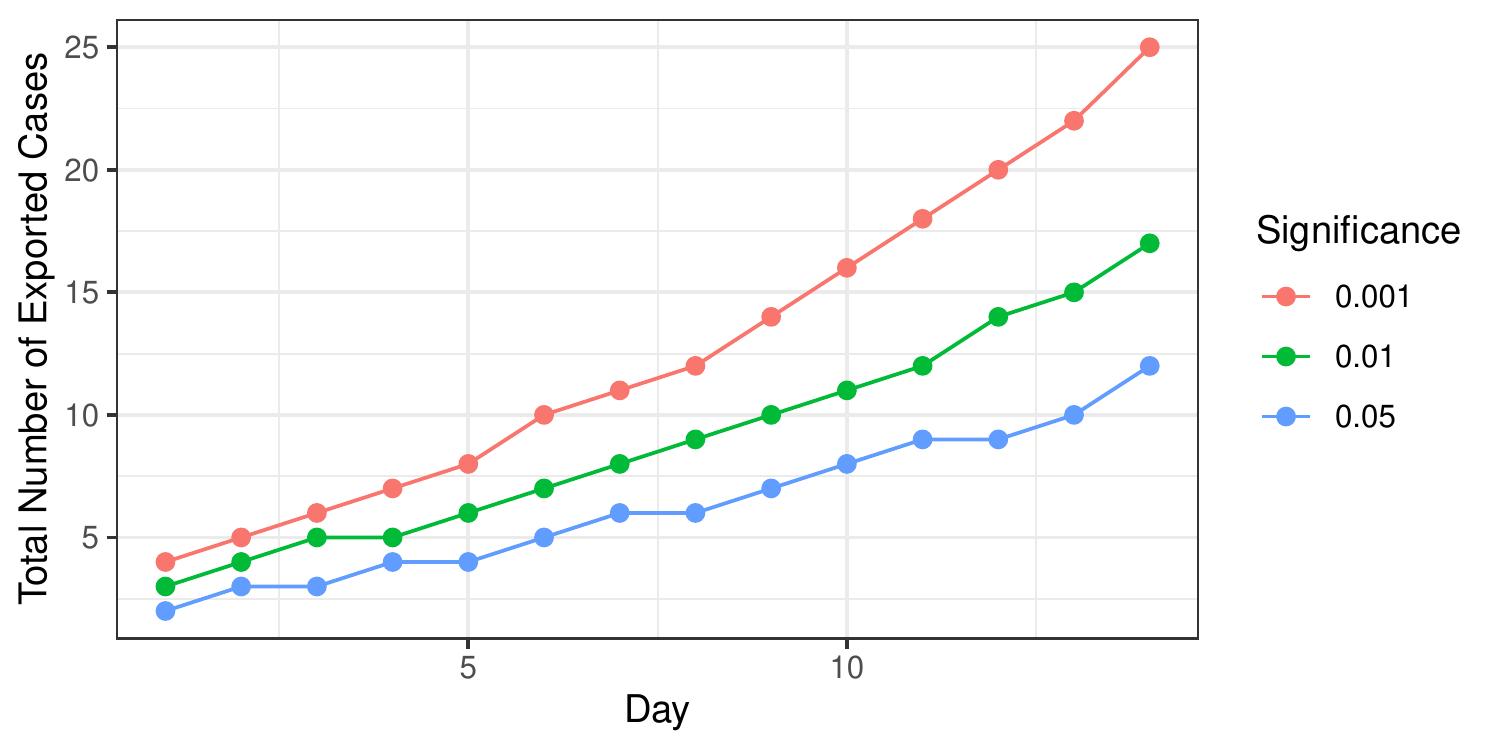}
\end{subfigure}
\caption{Hypothesis test of whether the exponential growth rate exceeds 0.1. The require cumulative total of traveler cases to reject the null hypothesis. The tests were conducted under three different significance levels: 0.05 (blue), 0.01 (green), and 0.001 (red). The x-axis indicates (a) the number of days since the initial local infection, (b) the number of days since the first exported case. The y-axis indicates the cumulative total of exported cases at
the day indicated by x-axis.}
\label{fig:testing}
\end{figure}

Figure \ref{fig:testing} illustrates the minimum required cumulative number of traveler cases for each day to detect $\beta_1>0.1$, in the case of known (top) and unknown (bottom) initial local infection date. The significance levels were set to be 0.05, 0.01 and 0.001. First of all, more traveler cases were needed to draw the conclusion at a higher significance level. The number of cases needed increased as the epidemic went into a later period. In an extreme case, if on the first day of the epidemic there was already an exported case, we would be highly confident that the local infection rate was very high. When making decisions to detect a future disease outbreak on a certain day, if we have relatively accurate information about how long it has been since the first local infection, we can use the top figure to compare the total number of exported cases up to this day. At a pre-specified significance level, say 5\%, if the total number has exceeded the number of the corresponding date in the figure, we would be able to tell that with 95\% confidence the local epidemic has an exponential growth rate bigger than 0.1. If we do not know the initial infection date, we would count how many days it has been since the first exported case, and compare with the bottom figure.   

A Shinny App for implementing the above hypothesis testing was provided in Appendix C, where users could specify the initial prevalence, the null hypothesis, the daily average of outbound traveler sizes, the number of simulations, and the significance level. With this tool, policy makers could adjust the parameters to fit their own country's situation, and quickly determine how severe their country's epidemic is based on the exported cases.

\section{Discussion}

We used the COVID-19 epidemic as an example to illustrate how the traveler case reports could be used to detect a disease outbreak at early stage. We found that the dates that our estimated indicators exceeded certain threshold (exponential growth rate were significantly above 0.1 (doubling time $<$ 7 days), and probability of COVID-19 patients exceeding hospital capacity $>$ 90\%) were all within the period that critical policies (city lock-downs and announcing national emergencies) were made based on domestic data. In general, using the traveler data was effective in detecting a disease outbreak in the travel origin.

From the data impact study, we found that knowing the actual arrival dates improved the estimates. In addition, more detailed travel history at sub-national level helped countries identify whether the outbreak was within a region or at the national level. For instance, without such information for patients with travel history to Egypt, we could not tell whether the initial epidemic was limited to tourism sites. Given the spreading speed of an emerging infection, it would be desirable to have early warning at sub-national levels. The proposed method can also detect the emerging disease outbreak at sub-national level if the sub-national surveillance system traces and reports the travel history of patients who have visited other sub-national areas. Implementation of such analysis requires travel size estimates across sub-national units in addition to the traveler case reports.

When comparing the estimates from the traveler data to the ones from domestic data, we found that the using only the traveler data had several limitations. One would need relatively accurate information about when the epidemic first started in the travel origins. A large estimate of exponential growth rate from the travel data could imply either a fast growth rate or a severe under-reporting of initial cases. More fundamentally, how well one can estimate the domestic epidemic using the traveler data depends on how representative the traveler samples are. As an extreme example, if the travelers were all young adults with good health, using the traveler data would've underestimated the epidemic in the general population. In our model, we introduced a non-zero intercept $\beta_0$ in the exponential growth curve, trying to correct the sampling bias. However, without further information, we could only use a non-informative prior on $\beta_0$. Studies that could potentially improve this sampling bias correction include, but are not limited to, better understanding the travelers' demographics, such as age, gender, sub-national region of residence, etc.

Finally, we provided an easy-to-use tool for policy makers to determine the growth rate of local epidemic based on exported cases. A user-friendly Shiny App was developed to accommodate flexible scenarios that fit different countries' situations. 

Rapid detection of early disease outbreak is crucial for government intervention and raising public awareness. The traveler case reports data seems to be a good addition to the domestic surveillance data by utilizing the diagnosis resources from all countries. We advocate that countries should work in a collaborative way, by sharing the traveler patients information about the travel dates and more detailed travel history at sub-national level, in a timely manner. Working together, we would strengthen the global infectious disease surveillance system, which is especially important in detecting disease outbreaks in countries where public health infrastructure is rudimentary or nonexistent.

\bibliographystyle{model1-num-names}
\bibliography{COVID19}

\begin{thebibliography}{17}
\expandafter\ifx\csname natexlab\endcsname\relax\def\natexlab#1{#1}\fi
\providecommand{\bibinfo}[2]{#2}
\ifx\xfnm\relax \def\xfnm[#1]{\unskip,\space#1}\fi
\bibitem[{{World Health Organization (WHO) Emergency
  Committee}(2020)}]{WHO2020Jan31}
\bibinfo{author}{{World Health Organization (WHO) Emergency Committee}},
\newblock \bibinfo{title}{Statement on the second meeting of the {International
  Health Regulations} (2005) emergency committee regarding the outbreak of
  novel coronavirus (2019-ncov)}  (\bibinfo{year}{2020}).
\bibitem[{Heymann et~al.(2001)Heymann, Rodier et~al.}]{Heymann2001hot}
\bibinfo{author}{D.~L. Heymann}, \bibinfo{author}{G.~R. Rodier}, et~al.,
\newblock \bibinfo{title}{Hot spots in a wired world: {WHO} surveillance of
  emerging and re-emerging infectious diseases},
\newblock \bibinfo{journal}{The Lancet Infectious Diseases} \bibinfo{volume}{1}
  (\bibinfo{year}{2001}) \bibinfo{pages}{345--353}.
\bibitem[{Morse(2007)}]{Morse2007global}
\bibinfo{author}{S.~S. Morse},
\newblock \bibinfo{title}{Global infectious disease surveillance and health
  intelligence},
\newblock \bibinfo{journal}{Health Affairs} \bibinfo{volume}{26}
  (\bibinfo{year}{2007}) \bibinfo{pages}{1069--1077}.
\bibitem[{Wilson(1995)}]{Wilson1995travel}
\bibinfo{author}{M.~E. Wilson},
\newblock \bibinfo{title}{Travel and the emergence of infectious diseases.},
\newblock \bibinfo{journal}{Emerging infectious diseases} \bibinfo{volume}{1}
  (\bibinfo{year}{1995}) \bibinfo{pages}{39}.
\bibitem[{Wu et~al.(2020)Wu, Leung, and Leung}]{Wu2020nowcasting}
\bibinfo{author}{J.~T. Wu}, \bibinfo{author}{K.~Leung}, \bibinfo{author}{G.~M.
  Leung},
\newblock \bibinfo{title}{Nowcasting and forecasting the potential domestic and
  international spread of the 2019-ncov outbreak originating in {Wuhan, China}:
  a modelling study},
\newblock \bibinfo{journal}{The Lancet} \bibinfo{volume}{395}
  (\bibinfo{year}{2020}) \bibinfo{pages}{689--697}.
\bibitem[{Imai et~al.(2020)Imai, Dorigatti, Cori, Donnelly, Riley, and
  Ferguson}]{Imai2020report2}
\bibinfo{author}{N.~Imai}, \bibinfo{author}{I.~Dorigatti},
  \bibinfo{author}{A.~Cori}, \bibinfo{author}{C.~Donnelly},
  \bibinfo{author}{S.~Riley}, \bibinfo{author}{N.~M. Ferguson},
\newblock \bibinfo{title}{Report 2: Estimating the potential total number of
  novel coronavirus cases in {Wuhan City, China}},
\newblock \bibinfo{journal}{Imperial College London}  (\bibinfo{year}{2020}).
\bibitem[{Chinazzi et~al.(2020)Chinazzi, Davis, Gioannini, Litvinova, Rossi,
  Xiong, Halloran, and Vespignani}]{Chinazzi2020preliminary}
\bibinfo{author}{M.~Chinazzi}, \bibinfo{author}{J.~T. Davis},
  \bibinfo{author}{C.~Gioannini}, \bibinfo{author}{M.~Litvinova},
  \bibinfo{author}{L.~Rossi}, \bibinfo{author}{X.~Xiong},
  \bibinfo{author}{M.~E. Halloran}, \bibinfo{author}{A.~Vespignani},
\newblock \bibinfo{title}{Preliminary assessment of the international spreading
  risk associated with the 2019 novel coronavirus (2019-ncov) outbreak in
  {Wuhan} city},
\newblock \bibinfo{journal}{Lab. Model. Biol. Soc.--Techn. Syst}
  (\bibinfo{year}{2020}).
\bibitem[{Xu et~al.(2020)Xu, Gutierrez, Mekaru, Sewalk, Goodwin, Loskill, Cohn,
  Hswen, Hill, Cobo et~al.}]{xu2020epidemiological}
\bibinfo{author}{B.~Xu}, \bibinfo{author}{B.~Gutierrez},
  \bibinfo{author}{S.~Mekaru}, \bibinfo{author}{K.~Sewalk},
  \bibinfo{author}{L.~Goodwin}, \bibinfo{author}{A.~Loskill},
  \bibinfo{author}{E.~L. Cohn}, \bibinfo{author}{Y.~Hswen},
  \bibinfo{author}{S.~C. Hill}, \bibinfo{author}{M.~M. Cobo}, et~al.,
\newblock \bibinfo{title}{Epidemiological data from the covid-19 outbreak,
  real-time case information},
\newblock \bibinfo{journal}{Scientific Data} \bibinfo{volume}{7}
  (\bibinfo{year}{2020}) \bibinfo{pages}{1--6}.
\bibitem[{Pan et~al.(2020)Pan, Liu, Wang, Guo, Hao, Wang, Huang, He, Yu, Lin
  et~al.}]{Pan2020association}
\bibinfo{author}{A.~Pan}, \bibinfo{author}{L.~Liu}, \bibinfo{author}{C.~Wang},
  \bibinfo{author}{H.~Guo}, \bibinfo{author}{X.~Hao},
  \bibinfo{author}{Q.~Wang}, \bibinfo{author}{J.~Huang},
  \bibinfo{author}{N.~He}, \bibinfo{author}{H.~Yu}, \bibinfo{author}{X.~Lin},
  et~al.,
\newblock \bibinfo{title}{Association of public health interventions with the
  epidemiology of the covid-19 outbreak in wuhan, china},
\newblock \bibinfo{journal}{The Journal of the American Medical Association}
  \bibinfo{volume}{323} (\bibinfo{year}{2020}) \bibinfo{pages}{1915--1923}.
\bibitem[{{The Center for Systems Science and Engineering (CSSE) at Johns
  Hopkins University}(2020)}]{JHUdata}
\bibinfo{author}{{The Center for Systems Science and Engineering (CSSE) at
  Johns Hopkins University}}, \bibinfo{title}{{(COVID-19)} data repository},
  \bibinfo{howpublished}{\url{https://github.com/CSSEGISandData/COVID-19}},
  \bibinfo{year}{2020}.
\bibitem[{Hao et~al.(2020)Hao, Cheng, Wu, Wu, Lin, and
  Wang}]{Hao2020reconstruction}
\bibinfo{author}{X.~Hao}, \bibinfo{author}{S.~Cheng}, \bibinfo{author}{D.~Wu},
  \bibinfo{author}{T.~Wu}, \bibinfo{author}{X.~Lin}, \bibinfo{author}{C.~Wang},
\newblock \bibinfo{title}{Reconstruction of the full transmission dynamics of
  covid-19 in wuhan},
\newblock \bibinfo{journal}{Nature}  (\bibinfo{year}{2020})
  \bibinfo{pages}{1--7}.
\bibitem[{Liang et~al.(2020)Liang, Guan, Li, Li, Liang, Zhao, Liu, Sang, Chen,
  Tang et~al.}]{liang2020clinical}
\bibinfo{author}{W.~Liang}, \bibinfo{author}{W.~Guan}, \bibinfo{author}{C.~Li},
  \bibinfo{author}{Y.~Li}, \bibinfo{author}{H.~Liang},
  \bibinfo{author}{Y.~Zhao}, \bibinfo{author}{X.~Liu},
  \bibinfo{author}{L.~Sang}, \bibinfo{author}{R.~Chen},
  \bibinfo{author}{C.~Tang}, et~al.,
\newblock \bibinfo{title}{Clinical characteristics and outcomes of hospitalised
  patients with {COVID}-19 treated in {Hubei} (epicenter) and outside {Hubei}
  (non-epicenter): A nationwide analysis of {China}},
\newblock \bibinfo{journal}{European Respiratory Journal}
  (\bibinfo{year}{2020}).
\bibitem[{{Xinhua Net}(2020)}]{xinhuanet}
\bibinfo{author}{{Xinhua Net}}, \bibinfo{title}{Wuhan: Three designated
  hospitals treat patients with new coronavirus infection and pneumonia
  confirmed for free treatment},
  \bibinfo{howpublished}{\url{http://www.xinhuanet.com/2020-01/21/c_1125491413.htm}},
  \bibinfo{year}{2020}.
\bibitem[{Thorarinsdottir et~al.(2013)Thorarinsdottir, Gneiting, and
  Gissibl}]{Thorarinsdottir2013using}
\bibinfo{author}{T.~L. Thorarinsdottir}, \bibinfo{author}{T.~Gneiting},
  \bibinfo{author}{N.~Gissibl},
\newblock \bibinfo{title}{Using proper divergence functions to evaluate climate
  models},
\newblock \bibinfo{journal}{SIAM/ASA Journal on Uncertainty Quantification}
  \bibinfo{volume}{1} (\bibinfo{year}{2013}) \bibinfo{pages}{522--534}.
\bibitem[{Parsons and Bao(2018)}]{Parsons2018value}
\bibinfo{author}{J.~Parsons}, \bibinfo{author}{L.~Bao},
\newblock \bibinfo{title}{The value of information in retrospect},
\newblock \bibinfo{journal}{arXiv preprint arXiv:1806.01458}
  (\bibinfo{year}{2018}).
\bibitem[{{World Health Organization} et~al.(2020)}]{world2020coronavirus}
\bibinfo{author}{{World Health Organization}}, et~al.,
  \bibinfo{title}{Coronavirus disease ({COVID}-2019) situation reports},
  \bibinfo{year}{2020}.
\bibitem[{Negida(2020)}]{Negida2020estimation}
\bibinfo{author}{A.~Negida},
\newblock \bibinfo{title}{Comments on the estimation of the covid-19 burden in
  {Egypt} through exported case detection},
\newblock \bibinfo{journal}{The Lancet Infectious Diseases}
  (\bibinfo{year}{2020}).

\end{thebibliography}







\end{document}


\begin{frontmatter}



\title{Appendix A of ``Detecting Disease Outbreak Using Travelers Data -- A Case Study of COVID-19 Epidemic"}


\author{Le Bao et. al.}

\address{Department of Statistics, Penn State University\\
323 Thomas Building, University Park, PA 16802}

\end{frontmatter}

\linenumbers
\section{Study periods and associated travel volumes}

Here we provide the reasons of choosing the start date of COVID-19 epidemic and the end date of outbound travelers for each city / country:
\begin{itemize}
\item Wuhan: The symptom onset day of the first known case in Wuhan was on Dec 1, 2019 \cite{huang2020clinical}. The lock-down of Wuhan, China was effective on Jan 23, 2020. The first exported case was reported on Jan 13, 2020.
\item Unites States: The first case was confirmed on Jan 20, 2020 (\cite{world2020coronavirus}). The first exported case was reported on March 6, 2020. The national emergency of United States was announced on Mar 13, 2020. 
\item Italy: The first two cases were confirmed on Jan 31, 2020 (\cite{world2020coronavirus}). The first exported case was reported on Feb 24, 2020. The initial lock-down of Lodi in Italy began on Feb 21, 2020. Therefore, we further extend 7 days after the initial lock-down for our study period in Italy to cover more infected cases travelled before Feb 21. In order to obtain an estimate of growth rate at the early stage of disease outbreak before the interventions came into effect, we excluded the confirmed cases travelled after the initial lockdown on Feb 21.
\item Iran: The first two deaths were announced on Feb 19, 2020 (\cite{world2020coronavirus}). We assumed that it took 20 days for cases from symptom onset to death, same as the first death case in Wuhan on Jan 9, 2020 with symptom onset on Dec 20, 2019. We focused on exported cases that traveled until Feb 27, 2020, 7 days after Feb 20, 2020, the first day of exported cases from Iran.
\item Egypt: The first case was confirmed on Feb 14, 2020 (\cite{world2020coronavirus}). The first exported case was reported on Feb 28, 2020. A lot more awareness of the COVID-19 outbreak were drawn on March 6, 2020 (\cite{tuite2020estimation}) as at least 14 foreign cases were associated with travel history in Egypt.
\end{itemize}

The outbound travel volume data for each city / country was obtained from the following sources:
\begin{itemize}
\item 2019 OAG data was used for Wuhan, China with adjusting the Hong Kong flight travel volume   \cite{Wu2020nowcasting}.
\item 2019 Q1 ICAO on flight origin and destination data (\url{https://data.icao.int/}) was used for United States.
\item Since a large proportion of international travelers took trains or buses for Italy, Iran and Egypt, 2018 UNWTO outbound tourism data \cite{UNWTO_outbound} was used for Italy, Iran and in 2015 for Egypt. We take into consideration that the decrease of travel volume due to COVID-19 during February (9\% decrease) and March 2020 (57\% decrease) according to UNWTO's 2020 tourism data (\cite{UNWTO_drop}).
\end{itemize}

Since the OAG and ICAO data only provide the flight travel volumes, we excluded the confirmed cases who did not travel by flight for data in Wuhan. We also excluded the confirmed cases in Mexico and Canada from United States due to their unknown travel transportation from United States to those destinations. Finally, two cases travelled from Italy via Hong Kong on Jan 31, 2020 confirmed by Taiwan on Feb 6, 2020 were believed to be infected on the Hong Kong flight (\cite{xinhuanet_taiwan}) and thus were excluded from the Italy data.

\section{Bayesian model details}

A weak prior was assigned to $\alpha$ implying that the prevalence ratio between the traveler population and the catchment population was 95\% likely to be in the range of $(0.1, 10)$: $\alpha & \sim Normal \big( 0,(\frac{\log(10)}{2})^2 \big)$. Non-informative prior was used for the exponential growth rate: $p(\beta_1) &\propto 1$. 

The time intervals between arrival and case confirmation only exhibit a small amount of variation within a country. So we use a weak prior for square root of dispersion parameter $\sqrt{1/\phi} \sim \text{half-Cauchy}(0,0.16)$ with 95\% probability falling into $(0,2)$ \cite{simpson2017penalising,gelman2006prior}. We put a weakly informative prior on $\lambda$, $\lambda \sim Gamma(0.6,0.1)$, that had a mean time interval 6 days and 95\% probability falling into $(0, 21)$ days. The gamma prior has mean and variance both $\lambda$.
\section{Other stuff}



\textcolor{red}{Here are results from existing literature:}
\begin{itemize}
    \item Lin, movement rate 0.05 and 0.08, $D_e=2.9$, $D_p=2.3$, $D_i=2.9$, removal rate $0.05+1/7.5=0.183$, $0.08+1/7.5=0.213$. transmission rate $b_{12}=1.31$. 
    Serial interval is $D_e+D_p/2+D_i/2 = 5.5$ days. Ascertainment rate $r=0.23$, and unascertainment case have lower transmissibility at rate $\alpha=0.55$. Exponential growth rate $log(1.31*(0.23+0.77*0.55)/0.183)/5.5=0.1581757$.
    \begin{itemize}
    \item[] $\log (\alpha \times b_{12} / (D_p^{-1}+n/N)) \times (D_p^{-1}+n/N)^{-1} +$
    \item[] $(1-r) \times \log (\alpha \times b_{12} / (D_i^{-1}+n/N)) \times (D_i^{-1}+n/N)^{-1}+$
    \item[] $r \times \log (b_{12} / (D_i^{-1}+D_q^{-1})) \times (D_i^{-1}+D_q^{-1})^{-1}$.
    \end{itemize}
    \item Nowcasting and forecasting the potential domestic and international spread of the 2019-nCoV outbreak originating in Wuhan, China: a modelling study \cite{Wu2020nowcasting}: data from Dec 31, 2019, to Jan 28, 2020 to infer the number of infections in Wuhan from Dec 1, 2019, to Jan 25, 2020.
    SEIR model with zoonotic force of infection equal to 86 cases per day before market closure on Jan 1, 2020, and equal to 0 thereafter.
    \textcolor{red}{serial interval = 8.4 days; incubation period = 6 days; infectious period = 8.4-6 = 2.4 days. Exponential growth rate = $log(2)/6.4 = 0.108$, doubling time = 6.4 days; $R_0=2.68$.}
    \item Early Transmission Dynamics in Wuhan, China, of Novel Coronavirus–Infected Pneumonia \cite{Li2020early}: data among the first 425 patients; cases with illness onset between December 10 and January 4. SEIR model with unknown zoonotic force.
    \textcolor{red}{serial interval = 7.5 days; incubation period = 5.2 days; infectious period = 7.5-5.2 = 2.1 days. Exponential growth rate = $0.10$, doubling time = 7.4 days; $R_0=2.2$.}
    \item An investigation of transmission control measures during the first 50 days of the COVID-19 epidemic in China \cite{Tianeabb6105}: By fitting an epidemic model to the time series of cases reported in each province (fig. S3), we estimate \textcolor{red}{$R_0=3.15$ prior to the implementation of the emergency response on 23 January.}
    \item Risk for Transportation of Coronavirus Disease from Wuhan to Other Cities in China \cite{Du2020risk}: By fitting our epidemiologic model to data on the first 19 cases reported outside of China, we estimate \textcolor{red}{an epidemic doubling time of 7.31 days. $R_0=(1+rL)\times(1+rD)$, where L is the length of incubation period and D is the length of infectious period. Number of initial cases in Wuhan on December 1, 2019: 7.78. Epidemic growth rate $r=0.095$, doubling time = 7.31, $R_0=1.90$.}
    \item High Contagiousness and Rapid Spread of Severe Acute Respiratory Syndrome Coronavirus 2 \cite{Sanche2020high}: \textcolor{red}{Results show that the doubling time early in the epidemic in Wuhan was 2.3–3.3 days. Assuming a serial interval of 6–9 days, $R_0=5.7$. Growth rate estimates varied from 0.21 to 0.3.
    Very comprehensive domestic cases and travel data are available in appendix table.}
    \item Real-Time Estimation of the Risk of Death from Novel Coronavirus (COVID-19) Infection: Inference Using Exported Cases \cite{Jung2020real}: we modeled epidemic growth either from a single index case with illness onset on 8 December 2019, or using the growth rate fitted along with the other parameters based on data from 20 exported cases reported by 24 January 2020. \textcolor{red}{$R_0=1+rS$ was estimated to be 2.10 for single index case and 3.19 for 20 exported cases, where r=0.15 or 0.29 is the estimated growth rate and S=7.5 is the mean serial interval.}
    \item Serial interval of novel coronavirus (COVID-19) infections \cite{Nishiura2020serial}: \textcolor{red}{serial interval = 4.0 days; limiting our data to only the most certain pairs, serial interval = 4.0 days.}
    \item Incubation Period and Other Epidemiological Characteristics of 2019 Novel Coronavirus Infections with Right Truncation: A Statistical Analysis of Publicly Available Case Data \cite{Linton2020incubation}: 210 cases with confirmed COVID-19 infection and diagnosis outside of Wuhan, China. \textcolor{red}{incubation period = 5.0 days.}
\end{itemize}

\bibliographystyle{model1-num-names}
\bibliography{COVID19.bib}


\begin{frontmatter}



\title{Appendix B of ``Detecting Disease Outbreak Using Travelers Data -- A Case Study of COVID-19 Epidemic"}

\author{Le Bao et. al.}

\address{Department of Statistics, Penn State University\\
323 Thomas Building, University Park, PA 16802}

\end{frontmatter}

\section{Traveler Case Report Data}

\subsection{Traveler Case Report from Wuhan, China during Dec 1, 2019 - Jan 23, 2020}

\begin{table}[!ht]
\begin{tabular}{lllll}
Destination   & Confirm Date & Travel Date & Transportation& Number of Cases \\
Japan         & 2020-01-15   & 2020-01-06  & Flight &1   \\
Singapore     & 2020-01-23   & 2020-01-20  & Flight &1   \\
South Korea   & 2020-01-20   & 2020-01-19  & Flight  &1   \\
South Korea   & 2020-01-23   & 2020-01-22  & Flight &1   \\
Taiwan        & 2020-01-21   & 2020-01-20  & Flight &1   \\
Thailand      & 2020-01-13   & 2020-01-08  & Flight &1   \\
Thailand      & 2020-01-17   & 2020-01-13  & Flight &1   \\
Thailand      & 2020-01-22   &             & Flight &1   \\
Thailand      & 2020-01-22   & 2020-01-19  & Flight  &1   \\
United States & 2020-01-21   & 2020-01-15  & Flight  &1   \\
Vietnam       & 2020-01-23   & 2020-01-13  & Flight &1   \\
Hong Kong\footnote[1]     & 2020-01-23   &             & Railway  &2   \\
Macau\footnote[1]         & 2020-01-21   &             & Railway &1   \\
Macau\footnote[1]        & 2020-01-22   &             & Bus &1   \\
\end{tabular}
\caption{Exported cases from Wuhan to destinations outside of Mainland China as of Jan 23, 2020}
\end{table}
\footnotetext[1]{These cases were not included in our analysis as their transportation to destinations was not by flight.}

\subsection{Traveler Case Report from Iran during Feb 19, 2020 - Feb 27, 2020}
\begin{table}[]
\begin{tabular}{llll}
Destination                 & Confirm Date & Travel Date & Number of Cases \\
Afghanistan          & 2020-02-24   &             & 1               \\
Bahrain              & 2020-02-24   & 2020-02-24  & 1               \\
Bahrain              & 2020-02-25   &             & 9               \\
Bahrain              & 2020-02-26   &             & 10              \\
Canada               & 2020-02-20   &             & 1               \\
Canada               & 2020-02-26   & 2020-02-15  & 1               \\
Canada               & 2020-02-27   & 2020-02-24  & 1               \\
Georgia              & 2020-02-26   & 2020-02-25  & 1               \\
Iraq                 & 2020-02-24   &             & 1               \\
Iraq                 & 2020-02-25   &             & 4               \\
Iraq                 & 2020-02-27   &             & 1               \\
Kuwait               & 2020-02-24   & 2020-02-24  & 2               \\
Kuwait               & 2020-02-24   &             & 3               \\
Kuwait               & 2020-02-25   &             & 4               \\
Kuwait               & 2020-02-26   &             & 17              \\
Kuwait               & 2020-02-27   &             & 18              \\
Lebanon              & 2020-02-21   & 2020-02-20  & 1               \\
Lebanon              & 2020-02-26   &             & 1               \\
Lebanon              & 2020-02-27   & 2020-02-24  & 1               \\
Norway               & 2020-02-27   &             & 1               \\
Oman                 & 2020-02-24   &             & 2               \\
Oman                 & 2020-02-25   &             & 2               \\
Oman                 & 2020-02-27   &             & 2               \\
Pakistan             & 2020-02-26   &             & 1               \\
Pakistan             & 2020-02-26   & 2020-02-20  & 1               \\
Sweden               & 2020-02-27   & 2020-02-20  & 1               \\
United Arab Emirates & 2020-02-22   &             & 2              
\end{tabular}
\caption{Exported cases from Iran as of Feb 27, 2020}
\end{table}

\subsection{Traveler Case Report from Italy during Jan 31, 2020 - Feb 28, 2020}

\begin{table}[]
\begin{tabular}{llll}
Destination    & Confirm Date & Travel Date              & Number of Cases \\
Algeria        & 2020-02-25   & 2020-02-17               & 1               \\
Austria        & 2020-02-26   &                          & 2               \\
Brazil         & 2020-02-26   & 2020-02-21               & 1               \\
Croatia        & 2020-02-26   &                          & 1               \\
Denmark        & 2020-02-28   & 2020-02-15               & 1               \\
Finland        & 2020-02-28   &                          & 1               \\
France         & 2020-02-26   &                          & 2               \\
France         & 2020-02-27   &                          & 1               \\
France         & 2020-02-28   &                          & 1               \\
Georgia        & 2020-02-28   &                          & 1               \\
Germany        & 2020-02-25   &                          & 1               \\
Germany        & 2020-02-28   &                          & 2               \\
Greece         & 2020-02-26   &                          & 1               \\
Greece         & 2020-02-28   &                          & 1               \\
Lithuanian     & 2020-02-27   &                          & 1               \\
Mexico         & 2020-02-28   & 2020-02-16 -- 2020-02-21 & 3               \\
Netherlands    & 2020-02-27   &                          & 1               \\
Netherlands    & 2020-02-28   &                          & 1               \\
Norway         & 2020-02-27   &                          & 2               \\
Norway         & 2020-02-28   &                          & 2               \\
Romania        & 2020-02-25   & 2020-02-05               & 1               \\
Romania        & 2020-02-28   &                          & 1               \\
Spain          & 2020-02-24   &                          & 1               \\
Spain          & 2020-02-25   & 2020-02-12 -- 2020-02-22 & 1               \\
Spain          & 2020-02-25   &                          & 3               \\
Spain          & 2020-02-26   &                          & 1               \\
Spain          & 2020-02-27   & 2020-02-12 -- 2020-02-22 & 1               \\
Spain          & 2020-02-27   & 2020-02-19 -- 2020-02-25 & 1               \\
Spain          & 2020-02-27   &                          & 6               \\
Sweden         & 2020-02-26   &                          & 1               \\
Sweden         & 2020-02-27   &                          & 1               \\
Sweden         & 2020-02-28   &                          & 1               \\
Switzerland    & 2020-02-25   &                          & 1               \\
Switzerland    & 2020-02-26   &                          & 1               \\
Switzerland    & 2020-02-27   & 2020-02-20               & 2               \\
Switzerland    & 2020-02-27   &                          & 1               \\
Switzerland    & 2020-02-28   &                          & 4               \\
United Kingdom & 2020-02-27   &                          & 1               \\
United Kingdom & 2020-02-28   &                          & 1              
\end{tabular}
\caption{Exported cases from Italy as of Feb 28, 2020}
\end{table}

\subsection{Traveler Case Report from Egypt during Feb 14, 2020 - Mar 06, 2020}

\begin{table}[]
\begin{tabular}{llll}
\hline
Destination    & Confirm Date & Travel Date             & Number of Cases \\ 
Canada         & 2020-03-03   & 2020-02-20              & 1               \\
Canada         & 2020-03-02   & 2020-02-20              & 2               \\
Canada\footnote[2]         & 2020-03-01   & 2020-02-20  & 1               \\
Canada\footnote[2]         & 2020-02-28   & 2020-02-20  & 1               \\
Canada         & 2020-03-06   &                         & 1               \\
Greece         & 2020-03-05   &                         & 1               \\
Greece         & 2020-03-04   &                         & 1               \\
United States\footnote[3] & 2020-03-01   &              & 2               \\
United States  & 2020-03-06   &                         & 1               \\
France\footnote[4]        & 2020-02-29   & 2020-02-05-- 2020-02-16 & 6     \\
France\footnote[4]        & 2020-03-05   & 2020-02-05-- 2020-02-16 & 5     \\
Taiwan\footnote[5]         & 2020-02-29   &             & 1               \\ 
\end{tabular}
\caption{Exported cases from Egypt as of Mar 6, 2020}
\end{table}
\footnotetext[2]{This two Canada cases traveled to Egypt together hence we only count them as one case with the earliest confirm date on Feb 28, 2020.}
\footnotetext[3]{This two cases were family and traveled to Egypt together hence we only count them as one.}
\footnotetext[4]{There were 11 cases in the same touring group from France to Egypt. We only count them as one case with the earliest confirm date on Feb 29, 2020.}
\footnotetext[5]{This case had been to Nile River cruise trip in Egypt hence was excluded from our study.}

\subsection{Traveler Case Report from United States during Jan 20, 2020 - Mar 13, 2020}

\begin{table}[]
\begin{tabular}{llll}
Destination  & Confirm Date & Travel Date & Number of Cases \\
Australia    & 2020-03-06   & 2020-02-29  & 1               \\
Australia    & 2020-03-09   & 2020-02-29  & 1               \\
Australia    & 2020-03-09   & 2020-03-06  & 1               \\
Australia    & 2020-03-10   & 2020-02-29  & 1               \\
Australia    & 2020-03-10   & 2020-03-08  & 1               \\
Australia    & 2020-03-10   & NA          & 2               \\
Australia    & 2020-03-11   & 2020-02-29  & 1               \\
Australia    & 2020-03-11   & 2020-03-07  & 2               \\
Australia    & 2020-03-11   & NA          & 1               \\
Australia    & 2020-03-12   & NA          & 3               \\
Australia    & 2020-03-13   & NA          & 4               \\
Brazil       & 2020-03-12   & 2020-03-06  & 1               \\
Brazil       & 2020-03-13   & NA          & 1               \\
Guyana       & 2020-03-11   & 2020-03-05  & 1               \\
Philippines  & 2020-03-10   & NA          & 1               \\
Philippines  & 2020-03-11   & NA          & 1               \\
South Africa & 2020-03-12   & NA          & 2              
\end{tabular}
\caption{Exported cases from United States as of Mar 13, 2020}
\end{table}

\bibliographystyle{model1-num-names}
\bibliography{COVID19.bib}